\documentclass[
a4paper,
reprint,
onecolumn,
aps,
pre,
amsmath,
superscriptaddress,
longbibliography
]{revtex4-2}
\usepackage{graphicx}
\usepackage{dcolumn}
\usepackage{bm}
\usepackage{color}
\usepackage{xcolor}
\usepackage{xfrac}
\usepackage{mathrsfs}
\usepackage{comment}
\usepackage{mathtools}
\usepackage{titlesec}
\usepackage{appendix}
\usepackage{amsfonts}

\renewcommand{\thesection}{\arabic{section}}
\renewcommand{\thesubsection}{\thesection.\arabic{subsection}}
\renewcommand{\thesubsubsection}{\thesubsection.\arabic{subsubsection}}

\def\br{{\bm{r}}}
\def\tbr{{\tilde{\bm{r}}}}
\def\Pe{{\rm{Pe}}}
\def\bu{{\bm{\hat{u}}}}
\def\bR{{\bm{\mathcal{R}}}}

\def\tt{{\tilde{t}}}

\newcommand{\as}[1]{\begingroup\color[rgb]{0,0.5,1}#1\endgroup}

\begin{document}
\title{Impact of chirality on active Brownian particle: Exact moments in two and three dimensions}
\author{Anweshika Pattanayak}
\email{ph16036@iisermohali.ac.in}
\affiliation{Department of Physical Sciences, Indian Institute of Science Education and Research Mohali, Sector 81, Knowledge City, S. A. S. Nagar, Manauli PO 140306, India}
\author{Amir Shee}
\email{amir.shee@northwestern.edu}
\affiliation{Northwestern Institute on Complex Systems and ESAM,
Northwestern University, Evanston, IL 60208, United States of America}
\author{Debasish Chaudhuri}
\email{debc@iopb.res.in}
\thanks{corresponding author}
\affiliation{Institute of Physics, Sachivalaya Marg, Bhubaneswar 751005, India.}
\affiliation{Homi Bhabha National Institute, Anushaktigar, Mumbai 400094, India}
\author{Abhishek Chaudhuri}
\email{abhishek@iisermohali.ac.in}
\thanks{corresponding author}
\affiliation{Department of Physical Sciences, Indian Institute of Science Education and Research Mohali, Sector 81, Knowledge City, S. A. S. Nagar, Manauli PO 140306, India}
\date{\today}
\begin{abstract} 
In this work, we investigate the effects of chirality, accounting for translational diffusion, on active Brownian particles in two and three dimensions. Despite the inherent complexity in solving the Fokker-Planck equation, we demonstrate a Laplace transform method for precisely calculating the temporal evolution of various dynamic moments. Our analysis yields explicit expressions for multiple moments, such as the second and fourth moments of displacement, revealing the impact of persistence and chirality. These moments exhibit oscillatory behaviour, and excess kurtosis indicates deviations from the Gaussian distribution during intermediate time intervals.
\end{abstract}

\maketitle


\section{Introduction}

 Active matter is driven out of equilibrium at the shortest scale of individual constituents, consuming and dissipating energy from the local environment or internal energy source. It breaks time-reversal symmetry to generate self-propulsion or stress on the surroundings. Examples of such systems abound in nature across a wide range of length scales, starting from molecular motors, cytoskeleton, individual cells and bacteria, tissues and organisms, and collective properties of organisms, e.g., bird flocks, fish schools, or human crowds~\cite{marchetti2013hydrodynamics, romanczuk2012active,vicsek2012collective, mirkovic2010fuel,schweitzer2003brownian}. Drawing inspiration from them, artificial active matters are designed, examples of which include Janus colloids utilizing phoretic motion, vibrated granular matter, and hexbugs~\cite{golestanian2007designing, PhysRevLett.112.068301,ruckner2007chemically, PhysRevLett.99.048102,ke2010motion}.

The motion of self-propelled agents is often described in terms of three related models: the active Brownian particles~(ABP), run-and-tumble particles~(RTP), and active Ornstein-Uhlenbeck process~(AOUP). Up to the second moment, their dynamics are equivalent and can easily be mapped from one to another. The generation of self-propulsion often utilizes a breaking of parity in the direction of motion, the heading direction, which undergoes either continuous~(ABP, active colloids) or discrete reorientation~(RTP, bacteria). In the active phoretic motion of colloids, such asymmetry is inherent to the design of the Janus colloids. In vibrated granular matter, the frictional asymmetry between the front and back leads to the generation of persistent motion, utilizing the vibration in the plane transverse to the motion. However, in general, the left-right parity symmetry around the heading direction can also be broken. This leads to the chirality of the agents, forcing them to turn in the broken symmetry direction while performing self-propulsion~\cite{liebchen2022chiral,lauga2006swimming,ledesma2012circle,su2013sperm, ghosh2009controlled, friedrich2009steering, namdeo2014numerical}. 
Chirality in active matter is observed in various natural systems, such as bacteria near walls and interfaces~\cite{diluzio2005escherichia,lauga2006swimming,di2011swimming}, sperm cells that swim helically~\cite{riedel2005self,friedrich2009steering}, and the formation of chiral FtsZ rings before bacterial cytokinesis~\cite{loose2014bacterial}. In synthetic systems, colloidal microswimmers with broken chiral symmetry~\cite{kummel2013circular,ten2014gravitaxis,shelke2019transition,zhang2020reconfigurable,alvarez2021reconfigurable}, motile droplets~\cite{kruger2016curling,lancia2019reorientation}, granular ellipsoids~\cite{barois2020sorting,arora2021emergent}, and cholesteric droplets~\cite{carenza2019rotation} also show active chiral motion. 
The analysis of the dynamics of one of the simplest active chiral agent models, chiral ABPs (cABP) is the focus of this paper.

Chirality in active systems led to several remarkable properties, e.g., odd viscosity, odd elasticity, and odd diffusivity~\cite{fruchart2023odd, scheibner2020odd, Hargus2021a, Soni2019}. The motion of a cABP has been analytically characterized in terms of its mean trajectory in two and three dimensions~\cite{kummel2013circular,van2008dynamics,wittkowski2012self}. Other studies on single cABPs have examined the effects of medium~\cite{sprenger2022active} and confinement~\cite{van2008dynamics,ao2015diffusion,caprini2019active,fazli2021active,murali2022geometric}. Recent analytic calculations of the dynamics of active chiral particles used intermediate scattering functions~\cite{sevilla2016diffusion, Kurzthaler2017a}. Here, we adopt a different approach. We extend a Laplace transform-based method applied to the Fokker-Planck equation that has been used recently to calculate all dynamical moments of ABPs with and without speed fluctuations and in the presence or absence of inertia~\cite{shee2020active,chaudhuri2021active,shee2022active,patel2023exact, Patel2024} to cABPs. 
The method was originally proposed to study the worm-like chain model of polymers~\cite{hermans1952statistics, daniels1952xxi}. As we show in this paper, with the help of explicit calculations in both two and three dimensions, this method provides a unified approach to calculating all dynamical moments of cABPs, reproducing known results, and providing closed-form analytic expressions for many others, including components of MSD in the heading direction and perpendicular to it, fourth moment of displacement, and excess kurtosis in three dimensions.    

The rest of the paper is organized as follows. We separately consider the Langevin dynamics and the Fokker-Planck equations for cABPs first in two dimensions (2d) and then in three dimensions (3d). All results are obtained separately, including the second and fourth moments of displacement variables, the moments of their components in directions parallel and perpendicular to the original heading direction, and excess kurtosis. Finally, we conclude with a brief discussion and outlook.

\section{Dynamics of a chiral active Brownian particle in two dimensions (2d)}
\numberwithin{equation}{subsection}

We first consider a chiral active Brownian particle (cABP) in two dimensions. The dynamics of the particle can be described by its position $\br$ and orientation $\bu$ of the heading direction of self-propulsion with $\bu \cdot \bu = 1$. In two dimensions, if $\phi$ is the angle made by the orientation vector with the $x-$ axis, then in Cartesian coordinates $\bu = u_x{\bm{\hat{x}}} + u_y{\bm{\hat{y}}}$ with $u_x = \cos\phi$ and $u_y = \sin\phi$. The overdamped dynamics of the particle with a self-propulsion speed $v_0$ and an angular velocity $\omega$ about the $z-$axis can then be written as:
\begin{align}
\begin{aligned}
\dot{\bm{r}} &= v_0\bu+\boldsymbol{\xi}_T(t)
\\
\dot{\phi} &= \omega+ \xi_\phi(t)
\end{aligned}
\label{langevin2d}
\end{align}
where the translational noise $\bm{\xi}_T$ and the rotational noise $\xi_\phi$ have zero mean and variance given by, $\langle\xi_{Ti}(t)\xi_{Tj}(t^\prime)\rangle =2D\delta_{ij}\delta(t-t^\prime)$ and $\langle\xi_\phi(t)\xi_\phi(t^\prime)\rangle =2D_r\delta(t-t^\prime)$. Here $D$ and $D_r$ are the translational and rotational diffusion coefficients, respectively. We can use $\tau_r=D_r^{-1}$ as a unit of time and $\ell=\sqrt{D/D_r}$ as a unit of length. It is useful to express the strength of activity in terms of a dimensionless P{\'e}clet number $\Pe=v_0/\sqrt{D D_r}$ and the strength of chirality in terms of  $\Omega=\omega / D_r$. In dimensionless form, the rescaled length and time scales are expressed as $\tilde{\br} = \br/l$ and $\tilde{t} = t/\tau_r$, respectively. 

The above equations can be numerically integrated using the Euler-Maruyama scheme. However, in the following, we first demonstrate analytical methods to obtain exact calculations of all relevant dynamical moments using the Fokker-Planck description of the above stochastic process.

\numberwithin{equation}{subsection} 
\subsection{Fokker-Planck Equation and Derivation of the Moment Generation Equation}
The probability distribution function $P(\br,\bu,t)$ of the cABP can be described by the following Fokker-Planck equation (see Appendix~\ref{app_FP_2d} for detailed derivation), 
\begin{align}
\partial_t P &=D\nabla^2 P+D_r\partial_\phi^2 P- v_0\hat{\textbf{u}}\cdot\nabla P-\partial_\phi(\omega  P)
\label{FPE}
\end{align}

Performing a Laplace transform $\tilde{P}(\br,\bu,s) = \int_0^\infty dt e^{-st} (\br,\bu,t)$, the Fokker-Planck equation can be expressed as,
\begin{align}
-P(\br,\bu,0)+s\tilde{P}(\br,\bu,s)=D\nabla^2\tilde{P} +D_r\partial_\phi^2\tilde{P} - v_0\bu\cdot\nabla\tilde{P} -\partial_\phi(\omega\tilde{P})
\label{LFPE}
\end{align}
where the initial condition at $t=0$ is set by $P(\br,\bu,0)=\delta(\br)\delta(\bu -\bu_0)$, without any loss of generality. This leads to the following equation
\begin{align}
-\langle \psi\rangle_0+s\langle\psi\rangle_s = v_0\langle\bu\cdot\nabla\psi\rangle_s+D\langle\nabla^2\psi\rangle_s+D_r\langle\partial_\phi^2\psi\rangle_s+\langle \omega\partial_\phi\psi\rangle_s
\label{ME}
\end{align}
for the mean of an arbitrary dynamical variable $\psi$ defined as $\langle\psi\rangle_s = \int d\br d\bu \psi(\br,\bu)\tilde{P}(\br,\bu,s)$ where the initial condition is given by  
$\langle \psi \rangle_0 = \int d\br d\bu\, \psi(\br,\bu)\, P(\br,\bu,0)$.
This equation is used to obtain all the dynamical moments in 2d.

\numberwithin{equation}{subsection}
\subsection{Orientation Autocorrelation}
We begin by considering $\psi(\br,\bu) = \bu$. Then, 
$\langle\psi\rangle_0 = \bu_0, \langle\nabla^2\psi\rangle_s = 0, \langle\bu\cdot\nabla\psi\rangle_s = 0$ and $\partial_\phi^2\psi=-\psi$. Note that $\langle \psi \rangle_0(= \bu_0) = u_{0x}{\bm{\hat{x}}} + u_{0y}{\bm{\hat{y}}} = \cos\phi_0{\bm{\hat{x}}} + \sin\phi_0{\bm{\hat{y}}}$, where the initial orientation is determined by $\phi_0$. In addition, $\partial_\phi u_x=-u_y $ and $\partial_\phi u_y=u_x$. Substituting these relations in Eq.~\ref{ME}, and solving for $\langle u_x \rangle_s$ and $\langle u_y \rangle_s$ we get:
\begin{align}
\begin{aligned}
\langle u_x \rangle_s&=\frac{(s+D_r)u_{0x}-\omega u_{0y}}{(s+D_r)^2+\omega^2}
\\
\langle u_y \rangle_s&=\frac{(s+D_r)u_{0y}+\omega u_{0x}}{(s+D_r)^2+\omega^2}
\end{aligned}
\label{oc1}
\end{align} 
Performing inverse Laplace transform on the above expressions, we get the time evolution of the components of the orientation vector, $\bu(t)$ dependent on the initial orientation $\bu_0$ as, 
\begin{align}
\begin{aligned}
 \langle u_x (t)\rangle &=e^{-D_r t}\left[u_{0x}\cos(\omega t)-u_{0y}\sin(\omega t)\right]=e^{-D_r t} \cos (\omega t +\phi_0)
\\
\langle u_y (t)\rangle &=e^{-D_r t}\left[u_{0y}\cos(\omega t)+u_{0x}\sin(\omega t)\right]=e^{-D_r t}  \sin (\omega t +\phi_0)
\end{aligned}
\end{align} 
The orientation correlation is then obtained as follows~\cite{caprini2019active, Caprini2023}:
\begin{align}
\langle\bu\cdot\bu_0\rangle=e^{-D_r t}\cos (\omega\,t)
\label{eq:orientation_autocorr_2d}
\end{align}
which shows decaying oscillations with a time scale $\tau_r = 1/D_r$ and a time period $2\pi/\omega$ set by the angular velocity ($\omega$) of the chiral ABP. In the absence of chiral rotation ($\omega=0$), equation~(\ref{eq:orientation_autocorr_2d}) simplifies to the orientation autocorrelation of ABPs: $\langle\bu\cdot\bu_0\rangle=e^{-D_r t}$~\cite{shee2020active}. On the other hand, in the absence of rotational noise ($D_r=0$), the cABP motion simplifies to a steady deterministic active chiral motion: $\langle\bu\cdot\bu_0\rangle=\cos(\omega t)$.
\numberwithin{equation}{subsection}
\subsection{Mean Displacement}
With $\psi = \br$ in Eq.~\ref{ME}, we get $\langle\br\rangle_s = (v_0/s)\bu$. Using the expressions for $\langle u_x\rangle_s$ and $\langle u_y\rangle_s$ from Eq.~\ref{oc1}, and then performing the inverse Laplace transform, we get the following expressions for the time evolution of the components of the displacement vector:
\begin{align}
\begin{aligned}
&\langle x(t)\rangle=\frac{v_0}{D_r^2+\omega ^2} \left[-D_r e^{-D_r t} \cos (\omega t + \phi_0) + D_r\cos(\phi_0)- \omega\sin(\phi_0) +\omega e^{-D_r t} \sin (\omega t + \phi_0)\right]
\\
&\langle y(t)\rangle=\frac{v_0}{D_r^2+\omega ^2} \left[-D_r e^{-D_r t} \sin (\omega t + \phi_0) + D_r\sin(\phi_0) + \omega\cos(\phi_0) - \omega e^{-D_r t} \cos (\omega t + \phi_0)\right].
\end{aligned}
\end{align}
We can also define the displacement components along and perpendicular to the initial orientation as $\br_\parallel = (\br\cdot\bu_0)\bu_0$ and $\br_\perp = \br - \br_\parallel$. This gives us,
\begin{align}
    \begin{aligned}
        \langle\br_{\parallel}\rangle & =  \frac{v_0}{D_r^2+\omega^2}\left[D_r- D_r e^{-D_r t} \cos(\omega t)+\omega e^{-D_r t}\sin(\omega t)\right]\bu_0
        \\
        \langle\br_{\perp}\rangle & = \frac{v_0   }{D_r^2+\omega ^2}\left[\omega -D_r e^{-D_r t} \sin (\omega t)-\omega e^{-D_r t} \cos (\omega t )\right] \bu_0^\perp\,,
    \end{aligned}
\end{align}
where $\bu_0^\perp \cdot \bu_0=0$.
%

\subsection{Position-orientation cross-correlation}
An important quantity that can distinguish the chirality of cABPs is the position-orientation cross-correlation. Setting $\psi = \br\cdot\bu$, we get the cross-correlations in the direction of orientation and perpendicular to the direction of orientation as:
\begin{align}
    & \langle\br \cdot\bu\rangle = \frac{v_0 e^{-D_r t} \left(-D_r \cos (\omega t )+D_r e^{D_r t}+\omega  \sin (\omega t )\right)}{D_r^2+\omega ^2}
    \\
  & \langle\br \cdot\bu^{\perp}\rangle = \frac{v_0 e^{-D_r t} \left(  -e^{D_r t}\omega+D_r \sin (\omega t  )+\omega  \cos (\omega t  )\right)}{D_r^2+\omega ^2}
\end{align}
$\langle\br \cdot\bu\rangle$ is symmetric under chirality reversal but $\langle\br \cdot\bu^{\perp}\rangle$ changes under chirality reversal. Thus, $\langle\br \cdot\bu^{\perp}\rangle$ can identify whether the rotation is clockwise or counterclockwise. 

\begin{figure}[t]
\centering
  \includegraphics[width=\linewidth]{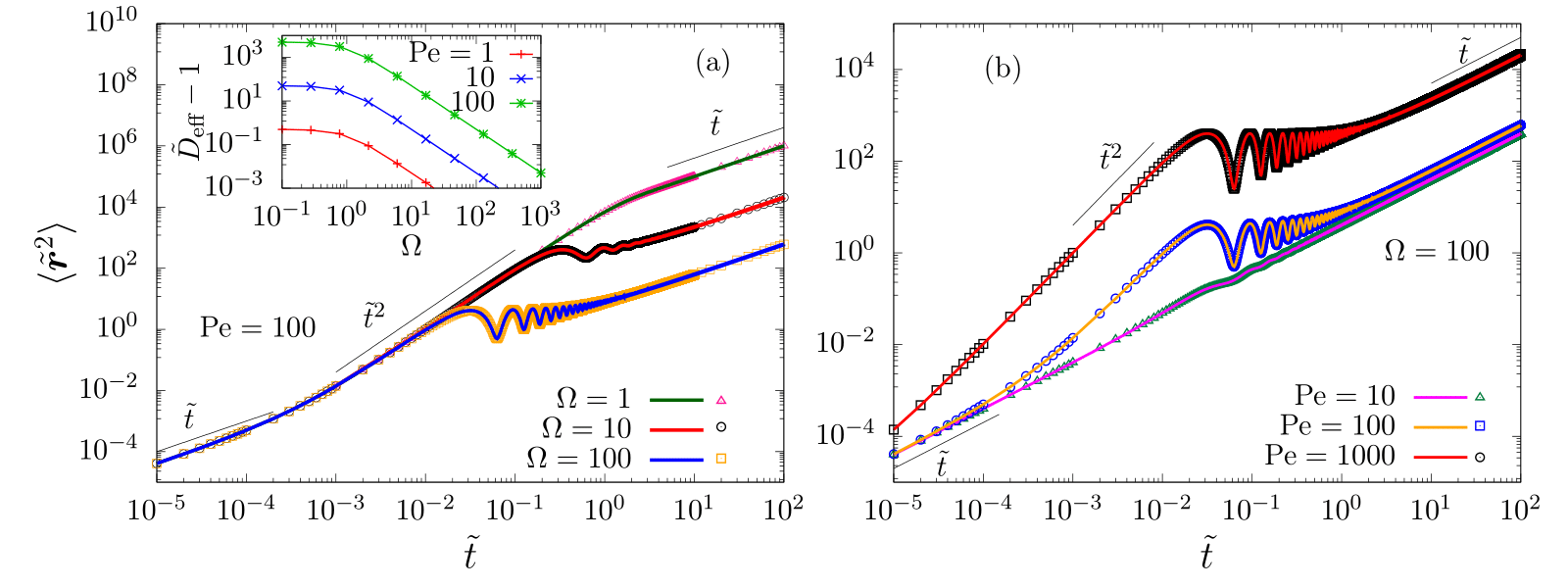}
  \caption{(a) MSD of 2d cABPs are plotted against time (in the unit of $D_r^{-1}$) for various $\Omega$ at a fixed  $\Pe=100$ indicated in the figure. The plots show oscillations at intermediate times at high $\Omega$.  (Inset) The active part of diffusivity shows saturation to different values of $\Pe$ in the ABP limit of $\Omega \rightarrow 0$, and $\Omega^{-2}$ scaling at large $\Omega$. (b) MSD for different $\Pe$ at a fixed $\Omega = 100$. For high $\Pe$($ = 1000$), the time scale for the initial diffusive to ballistic crossover ($\tt_I \sim 1/\Pe^2$) shifts to very low $\tt$. In both (a) and (b), the solid lines are plots of analytical expressions, and the points are obtained from simulations.}
  \label{msdPe2d}
\end{figure}
 
\numberwithin{equation}{subsection}
\subsection{Mean Squared Displacement (MSD)} \label{sectionMSD2d}
With $\psi = \br^2$, we get from Eq.~\ref{ME}, $s\langle \br^2\rangle_s=4D\langle 1\rangle_s+2v_0\langle \br\cdot\bu\rangle_s$ with $\langle 1\rangle_s = 1/s$. To get $\langle \br\cdot\bu\rangle_s$, we next set $\psi = \br\cdot\bu$. Again, substituting in Eq.~\ref{ME} and using $\nabla\psi = \bu$ and $\partial_\phi^2\psi = -2\psi$, we get $(s+D_r)\langle \br\cdot\bu\rangle_s=v_0/s + \omega\langle \partial_\phi(\textbf{r}\cdot\hat{\textbf{u}})\rangle_s$. To evaluate this, we set $\psi=\partial_\phi(\br\cdot\bu)=-(xu_y-yu_x)$, which gives:
\begin{align}
\begin{aligned}
\nabla\partial_\phi(\br\cdot\bu)&=\partial_\phi\nabla(\br\cdot\bu)=\partial_\phi(\bu)
\\
v_0\bu\cdot\nabla\partial_\phi(\br\cdot\bu)&=0
\\
\partial_\phi^2\partial_\phi(\br\cdot\bu)&=-\partial_\phi(\br\cdot\bu)
\\
\partial_\phi\psi=\partial_\phi^2(\br\cdot\bu)&=-(xu_x+yu_y)=-\br\cdot\bu
\end{aligned}
\end{align}
%

Hence, from Eq.~\ref{ME}, we get $(s+D_r)\langle\partial_\phi(\br\cdot\bu)\rangle_s=-\omega\langle\br\cdot\bu\rangle_s$ which on substitution gives 
\begin{align}
    \begin{aligned}
        \langle\br\cdot\bu\rangle_s = \frac{v_0(s+D_r)}{(s(s+D_r)^2+\omega^2)}.
    \end{aligned}
\end{align}
We can now evaluate the expression for $\langle \br^2\rangle_s$, and taking the inverse Laplace transform, we get the following expression for the mean square displacement~(MSD) 
\begin{align}
\begin{aligned}
\langle \br^2(t)\rangle=&\left(4D + \frac{2D_r v_0^2}{D_r^2+\omega ^2}\right)t \\
+&
\frac{2 v_0^2}{\left(D_r^2+\omega ^2\right)^2} \left[(\omega^2 - D_r^2)(1 - e^{-D_r t}\cos (\omega t)) - 2 D_r e^{-D_r t}\omega  \sin (\omega t)\right]
\label{msd2d}
\end{aligned}
\end{align}
Note that this expression is independent of the initial orientation.
The MSD shows an asymptotic diffusive scaling in $t\to\infty$ limit,  $\langle {\bf r}^2(t)\rangle = 4 D_{\rm eff} t$, with effective diffusion coefficient
\begin{align}
    D_{\rm eff}= D + \frac{v_0^2}{2 D_r} \frac{1}{1+\Omega^2}.
    \label{deff2d}
\end{align} 
Note that in the limit of vanishing chirality $\Omega \to 0$, the above expression reduces to the known behaviour of simple ABPs. Using $Pe$, the equation can be expressed as:
\begin{align}
\tilde{D}_{\rm eff} = \frac{D_{\rm eff}}{D} = 1+ \frac{\text{Pe}^2}{2} \frac{1}{(1 + \Omega^2)} \end{align}
This effective diffusion coefficient is plotted versus $\Omega$ for various $\text{Pe}$ in Fig.~\ref{msdPe2d}(a, Inset).

{\bf Dimensionless form:} In dimensionless form, the MSD can be expressed as:
\begin{align}
    \begin{aligned}
        \langle \tilde{\br}^2(\tilde t)\rangle &= \left(4 + \frac{2 \text{Pe}^2}{1 + \Omega ^2}\right) \tilde{t} \\
        &+ \frac{2\Pe^2}{(1 + \Omega^2)^2} \left[ (\Omega^2 - 1)(1 - e^{-\tt}\cos(\Omega\tt)) - 2e^{-\tt}\Omega\sin(\Omega\tt)\right] 
    \end{aligned}
\end{align}
This expression agrees with the results obtained in Ref.~\cite{Kurzthaler2017a}. Dimensionless MSD is plotted for different $\Pe$ and $\Omega$ in Fig.~\ref{msdPe2d}.  
At short times, expanding the MSD around $\tt = 0$ we find, 
\begin{align}
    \lim_{\tt\to 0} \langle \tbr^2(t)\rangle & = 4 \tt+\Pe^2t^2-\frac{\Pe^2}{3} \tt^3+\frac{\Pe^2}{12}  \left(1-\Omega ^2\right) \tt^4+O\left(\tt^5\right)\, .
\end{align}
Thus MSD exhibits equilibrium diffusion due to the heat bath $\langle \tbr^2\rangle=4\tt$ at shortest times, crossing over to active ballistic motion $\langle \tbr^2\rangle\simeq \text{Pe}^2\tt^2$, at $\tt_{I}=4/\text{Pe}^2$ (Fig.~\ref{msdPe2d}). The influence of deterministic chiral rotation on the MSD manifests in the coefficient of $\tt^4$. 



\par
Although the mean square displacement does not show any dependence on the initial orientation of cABPs, this dependence can be observed in the displacement correlations $\langle x^2\rangle$, $\langle y^2\rangle$, and $\langle xy\rangle$. To see this, we set $\psi=x^2$, $xy$, and $y^2$. The Laplace-transformed forms are the following:
\begin{align}
    \begin{aligned}
     \langle x^2\rangle_s &=\frac{2D}{s^2}
     \\ &-\frac{2 v_0^2\left[\omega ^2 \left(2 D_r+ 3s-2 s u_{x0}^2\right)+(D_r+s) (4 D_r+s) \left(2 D_r+s u_{x0}^2\right)-3 s u_{x0} u_{y0} \omega  (2 D_r+s)\right]}{s^2\left((D_r+s)^2+\omega ^2\right) \left((4 D_r+s)^2+4 \omega ^2\right)}
     \\
     \langle y^2\rangle_s &=\frac{2D}{s^2}
     \\ &-\frac{2 v_0^2\left[\omega ^2 \left(2 D_r+2 s u_{x0}^2+s\right)+(D_r+s) (4 D_r+s) \left(2 D_r+s u_{y0}^2\right)+3 s u_{x0} u_{y0} \omega  (2 D_r+s)\right]}{s^2\left((D_r+s)^2+\omega ^2\right) \left((4 D_r+s)^2+4 \omega ^2\right)}
     \\
     \langle xy\rangle_s &=\frac{v_0^2 \left(3 \left(2 u_{x0}^2-1\right) \omega  (2 D_r+s)+2 u_{x0} u_{y0} (D_r+s) (4 D_r+s)-4 u_{x0} u_{y0} \omega ^2\right)}{s \left((D_r+s)^2+\omega ^2\right) \left((4 D_r+s)^2+4 \omega ^2\right)}
    \end{aligned}
    \label{xxyycorr}
\end{align}
An average over all possible initial orientations will lead to $\langle u_{x0}^2\rangle = \langle u_{y0}^2\rangle = 1/2$ and $\langle u_{x0}u_{y0}\rangle = 0$. This gives $\langle xy\rangle_s=0$ and 
\begin{align}
    \langle x^2\rangle_s= \langle y^2\rangle_s=\frac{\left[(D_r+s) \left(2 D (D_r+s)+v_0^2\right)+2 D \omega ^2\right]}{\left[ s^2 \left((D_r+s)^2+\omega ^2\right)\right]}
\end{align}
One can easily find the exact expression of $\langle x^2(t)\rangle$, $\langle y^2(t)\rangle$, and  $\langle xy(t)\rangle$ by taking the inverse Laplace transformation of the above expressions. It is evident that when averaged over all possible initial orientations, the $x-y$ symmetry can be restored. However, as seen from Eq.~\ref{xxyycorr}, the correlations do depend on the initial orientation $\phi_0$, and, in general, the $x-y$ symmetry is broken. 

\subsubsection{MSD along and perpendicular to the initial orientation}
Without loss of generality, we choose the initial orientation along the $x-$ direction, $\bu_0=\hat{x}$. Then, with $u_{x0}=1$ and $u_{y0}=0$, we get:
\begin{align}
\begin{aligned}
    \langle \br_{\parallel}^2(t)\rangle &= \left(\frac{v_0^2 D_r}{D_r^2+\omega ^2}+2 D\right)t+ \frac{v_0^2 \left(7 \omega ^2 D_r^2-6 D_r^4+\omega ^4\right)}{2 \left(D_r^2+\omega ^2\right){}^2 \left(4 D_r^2+\omega ^2\right)}
    \\& -\frac{v_0^2 e^{-4 t D_r} \left(\left(\omega ^2-6 D_r^2\right) \cos (2 \omega t )+5 \omega  D_r \sin (2 \omega t )\right)}{2 \left(4 D_r^2+\omega ^2\right) \left(9 D_r^2+\omega ^2\right)}
    \\& + \frac{2 v_0^2 D_r e^{-t D_r} \left(\omega  \left(\omega ^2-7 D_r^2\right) \sin (\omega t )+D_r \left(3 D_r^2-5 \omega ^2\right) \cos (\omega t )\right)}{\left(D_r^2+\omega ^2\right){}^2 \left(9 D_r^2+\omega ^2\right)}
\end{aligned}
\end{align}
For $t \rightarrow 0$, 
$$ \langle \br_{\parallel}^2(t)\rangle = 2 D t+t^2 v_0^2-D_r t^3 v_0^2+\frac{1}{12} t^4 v_0^2 \left(11 D_r^2-4 \omega ^2\right)+O\left(t^5\right).$$
In the $t \to \infty$ limit, the effect of persistence disappears. The asymptotic behavior is diffusive where the diffusivity is governed by $D_{\rm eff}$ as obtained in Eq.~\ref{deff2d}.\\
We can similarly calculate MSD normal to the direction of the initial orientation:
\begin{align}
   \begin{aligned}
    \langle \br_{\perp}^2(t)\rangle &= t\left(\frac{v_0^2 D_r}{D_r^2+\omega ^2}+2 D\right)+ \frac{v_0^2 \left(5 \omega ^2 D_r^2-10 D_r^4+3 \omega ^4\right)}{2 \left(D_r^2+\omega ^2\right){}^2 \left(4 D_r^2+\omega ^2\right)}
    \\ & +\frac{v_0^2 e^{-4 t D_r} \left(\left(\omega ^2-6 D_r^2\right) \cos (2 \omega t )+5 \omega  D_r \sin (2 \omega t )\right)}{2 \left(4 D_r^2+\omega ^2\right) \left(9 D_r^2+\omega ^2\right)}
    \\ & -\frac{2 v_0^2 e^{-t D_r} \left(\omega  D_r \left(11 D_r^2+3 \omega ^2\right) \sin (\omega t )+\left(3 \omega ^2 D_r^2-6 D_r^4+\omega ^4\right) \cos (\omega t )\right)}{\left(D_r^2+\omega ^2\right){}^2 \left(9 D_r^2+\omega ^2\right)}
    \end{aligned}
\end{align}
For $t \rightarrow 0$, 
$$\langle \br_{\perp}^2(t)\rangle = 2 D t+\frac{2}{3} D_r t^3 v_0^2+\frac{1}{12} t^4 v_0^2 \left(3 \omega ^2-10 D_r^2\right)+O\left(t^5\right).$$
Like the parallel component, in the limit $t \to \infty$, as the effect of persistence disappears, the asymptotic behavior is diffusive where the diffusivity is governed by $D_{\rm eff}$.

\begin{figure}[t]
\centering
  \includegraphics[width=\linewidth]{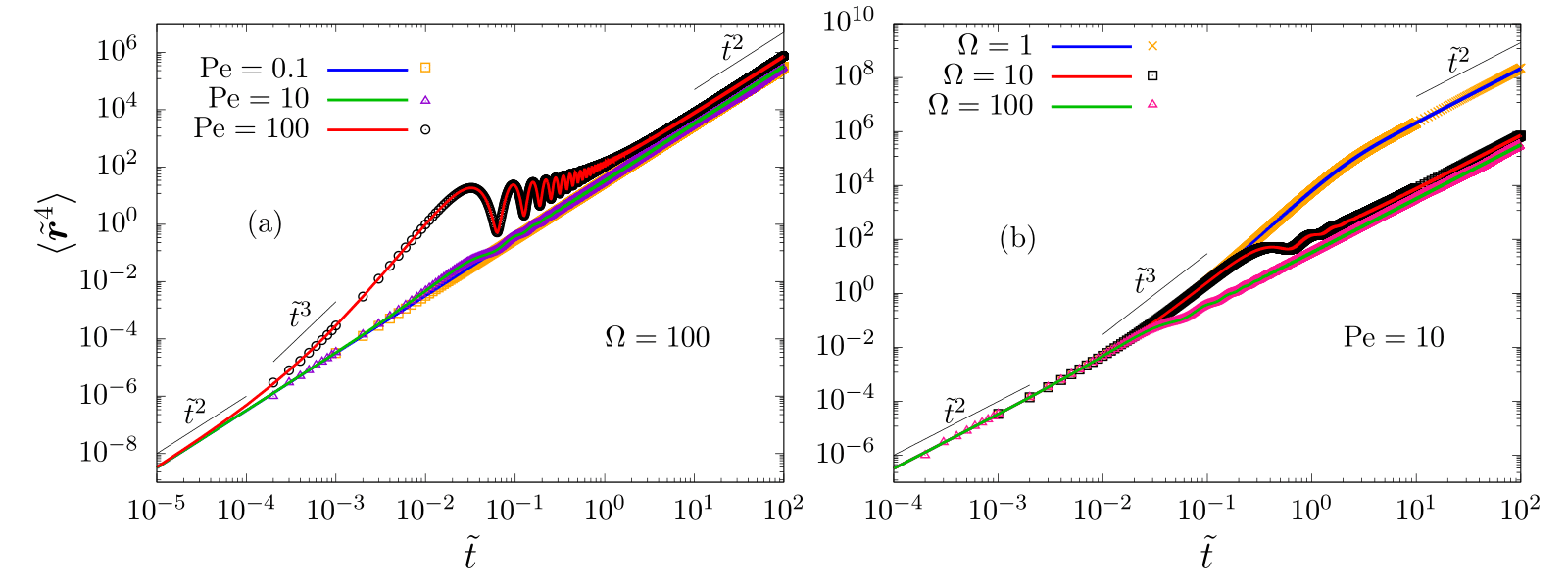}
  \caption{The fourth moments of the displacement of 2d cABPs are plotted against time (in the unit of $D_r^{-1}$) for (a) various $\Pe$ at a fixed $\Omega = 100$ and (b) various $\Omega$ at a fixed $\Pe$. Multiple crossovers are observed from $~\tt^2$ to $~\tt^3$. The solid lines denote analytical expressions in both (a) and (b), and the points are obtained from simulations.}
  \label{r4Pe2d}
\end{figure}

\numberwithin{equation}{subsection}
\subsection{Fourth Moment}
The general method of calculations from the Laplace transformed Fokker Planck equation for $\psi$ in Eq.~\ref{ME}, allows a straightforward albeit cumbersome calculation of the higher order moments. The fourth moment of displacement can be calculated by setting $\psi = \br^4$ in Eq.~\ref{ME}, to give: $s\langle \br^{4}\rangle_s=16 D\langle \br^{2}\rangle_s+4 v_{0}\langle \br^{2} \br\cdot\bu \rangle_{s}$. 
\subsubsection{Calculation of $\left\langle \br^{2} \br\cdot\bu\right\rangle_{s}$}
Substitution of $\psi = \br^2\br\cdot\bu$ in Eq.~\ref{ME} gives:
\begin{align}
\begin{aligned}
\left[s+D_r+\frac{\omega^{2}}{s+D_{r}}\right]\langle \br^2 \br\cdot\bu\rangle_s &=8 D\left[\langle\br\cdot\bu\rangle_{s}+\frac{\omega\langle \partial_\phi(\br\cdot\bu)\rangle_{s}}{s+D_{r}}\right]
\\&+ v_{0}\left[2 \left\langle(\br\cdot\bu)^{2}\right\rangle_{s} 
+\frac{\omega\langle \partial_\phi(\br\cdot\bu)^{2}\rangle_{s}}{s+D_{r}}+\langle \br^{2}\rangle_{s}\right]
\end{aligned}
\end{align}
\subsubsection{Calculation of $\left\langle(\br\cdot\bu)^{2}\right\rangle_{s}$}
Substitution of $\psi = (\br\cdot\bu)^2$ in Eq.~\ref{ME} gives:
\begin{align}
\begin{aligned}
\left[s+4 D_{r}+\frac{4 \omega^{2}}{s+4 D r}\right]\left\langle(\br\cdot\bu)^{2}\right\rangle_{s}&= \frac{2 D}{s}+2 D_{r}\left\langle \br^{2}\right\rangle_{s}+2 v_{0}\langle\br\cdot\bu\rangle_{s} \\& +\frac{2 \omega v_{0}\left\langle \partial_\phi(\br\cdot\bu)\right\rangle_{s}}{s+4 D_{r}} + \frac{2\omega^{2} \left\langle \br^{2}\right\rangle_{s}}{s+4 D_{r}} 
\end{aligned}
\end{align}
and similarly,
\begin{align}
\begin{aligned} 
{\left(s+4 D_{r}\right)\left\langle \partial_\phi(\br\cdot\bu)^{2}\right\rangle_{s} } &=2 v_{0}\left\langle \partial_\phi(\br\cdot\bu)\right\rangle_{s}-4 \omega\left\langle(\br\cdot\bu)^{2}\right\rangle_{z} +2 \omega\left\langle \br^{2}\right\rangle_{s}
\end{aligned}
\end{align}
Using these relations, we can write the final expression for $\langle \br^4\rangle$ in dimensionless form as 
\begin{align*}
	\left\langle \tilde{\br}^4(t)\right\rangle&=\frac{8 \left(\text{Pe}^2+2 \Omega ^2+2\right)^2}{\left(\Omega ^2+1\right)^2}\tilde{t}^2
	+\frac{8 \text{Pe}^2 \left[\text{Pe}^2 \left(2 \Omega ^4+11 \Omega ^2-15\right)+4 \left(\Omega ^6+4 \Omega ^4-\Omega ^2-4\right)\right]}{\left(\Omega ^2+1\right)^3 \left(\omega ^2+4\right)}\tilde{t} 
	\\
 &-\frac{16 \text{Pe}^2  \left[\text{Pe}^2 \left(\Omega ^4-12 \Omega ^2+3\right)-2 \left(\Omega ^2+9\right)+2 \left(\Omega ^2+9\right) \Omega ^4\right] }{\left(\Omega ^2+1\right)^3 \left(\Omega ^2+9\right)}\tilde{t} e^{-\tilde{t}} \cos (\Omega \tilde{t} )
 \\&-\frac{16 \text{Pe}^2  \left[\text{Pe}^2 \left((6 \tilde{t}-1) \Omega ^4+2 (6-5 \tilde{t}) \Omega ^2-3\right)+2 \left(\Omega ^2+1\right) \left(\Omega ^2+9\right) \left((2 \tilde{t}-1) \Omega ^2+1\right)\right] }{\Omega  \left(\Omega ^2+1\right)^3 \left(\Omega ^2+9\right)}e^{-\tilde{t}}\sin (\Omega \tilde{t})
	\\&+\frac{6 \text{Pe}^4 \left(\Omega ^8+9 \Omega ^6-61 \Omega ^4-241 \Omega ^2+116\right)}{\left(\Omega ^2+1\right)^4 \left(\Omega ^2+4\right)^2}
 \\&-\frac{8 \text{Pe}^4  \left(\Omega ^2-7\right) \left(\Omega ^6+11 \Omega ^4+139 \Omega ^2-63\right) }{\left(\Omega ^2+1\right)^4 \left(\Omega ^2+9\right)^2}e^{-\tilde{t}}\cos (\Omega \tilde{t})
	\\
	&-\frac{8 \text{Pe}^2  \left[\text{Pe}^2 \left(7 \Omega ^8+36 \Omega ^6+314 \Omega ^4-1116 \Omega ^2+135\right)+4 \left(\Omega ^2-1\right) \left(\Omega ^4+10 \Omega ^2+9\right)^2\right] }{\Omega  \left(\Omega ^2+1\right)^4 \left(\Omega ^2+9\right)^2}e^{-\tilde{t}}\sin (\Omega \tilde{t})
	\\
	&+ \frac{2 \text{Pe}^4  \left(10 \Omega  \left(\Omega ^2-6\right) \sin (2 \Omega \tilde{t} )+\left(\Omega ^4-37 \Omega ^2+36\right) \cos (2\Omega \tilde{t})\right)}{\left(\Omega ^2+4\right)^2 \left(\Omega ^2+9\right)^2} e^{-4\tilde{t}}
\end{align*}
Note that this expression is independent of the initial orientation. 
In the limit of vanishing chiral rotation $\Omega=0$, the above expression reduces to the known result for simple ABP derived earlier~\cite{shee2020active}.
In Fig.~\ref{r4Pe2d}, we plot $\langle \tbr^4 \rangle$ for cABP, for different $\Pe$ and $\Omega$ and compare the analytical results with numerical simulations. They show an exact match for all the different parameter values. 

In order to understand the dynamics at short times, we expand the expression as $\tt \to 0$, which yields
\begin{align}
    \lim_{\tt \to 0} \langle\tbr^4\rangle &=32 \tt^2+16 \text{Pe}^2 \tt^3+\frac{\text{Pe}^2}{3} \left(3 \text{Pe}^2-16\right) \tt^4-\frac{2\text{Pe}^2}{3} \left(\text{Pe}^2+2 \Omega^2-2\right)\tt^5+\mathcal{O}\left(\tt^6\right)
\end{align}
In the regime of short times, the fourth order moment exhibits $\langle\tbr^4\rangle=32\tt^2$, dominated by thermal fluctuations, which then crosses over to $\langle\tbr^4\rangle\simeq 16 \text{Pe}^2 \tt^3$, dominated by high activity, at $\tt_{I}=2/\text{Pe}^2$~(Fig.~\ref{r4Pe2d}). Note that this crossover is independent of $\Omega$ and is common to cABP and ABP. The chiral rotation manifests itself only in the coefficient of the further higher-order term $\tt^5$, suggesting that chirality influences the dynamics only at relatively longer time regimes. 
The late-time behavior of the fourth-order moment can be highlighted by focusing on the $\tt\to\infty$ limit,
\begin{align}
    \lim_{\tt\to\infty} \langle\tbr^4\rangle & = \frac{8 \left(2+\text{Pe}^2+2 \Omega^2\right)^2}{\left(1+\Omega ^2\right)^2}\tt^2 \, ,
\end{align}
showing the $\langle\tbr^4\rangle\sim \tt^2$ scaling as observed in Fig.~\ref{r4Pe2d}, where the coefficient depends on both activity $\text{Pe}$ and chirality $\Omega$.

\subsection{Excess Kurtosis: Deviation from Gaussian} 
If the displacement variable were a Gaussian process, its fourth moment can be expressed as $\tilde{\mu}_4=\langle \delta \tbr^2\rangle^2 + 2\langle \delta \tilde{r}_i \delta \tilde{r}_j \rangle^2 + 2\langle \delta \tbr^2\rangle\langle \tilde{r}\rangle^2
+ 4\langle \tilde{r}_i\rangle\langle \tilde{r}_j \rangle\langle \delta \tilde{r}_i\delta \tilde{r}_j \rangle + \langle \tilde{r}\rangle ^4$. This is the case for the chiral active Ornsetein-Uhlenbeck process (cAOUP)~\cite{caprini2019active, Caprini2023}. One can use this to extract the departure from the Gaussianity in terms of an excess kurtosis in dimensionless form 
\begin{equation}
    \tilde{\cal{K}} = \frac{\langle{\tbr^4}\rangle}{\mu_4} - 1\, .
\end{equation} 
If we average over the initial orientation of the particle, all the first moments vanish. The simplified expression of $\tilde{\mu}_4$ becomes:
\begin{align}
    \tilde{\mu}_4=\langle  \tbr^2\rangle^2 + 2\langle  \tilde{r}_i  \tilde{r}_j \rangle^2\, . 
    \label{mu4_avg}
\end{align}

\begin{figure}
    \centering
    \includegraphics[width=\linewidth]{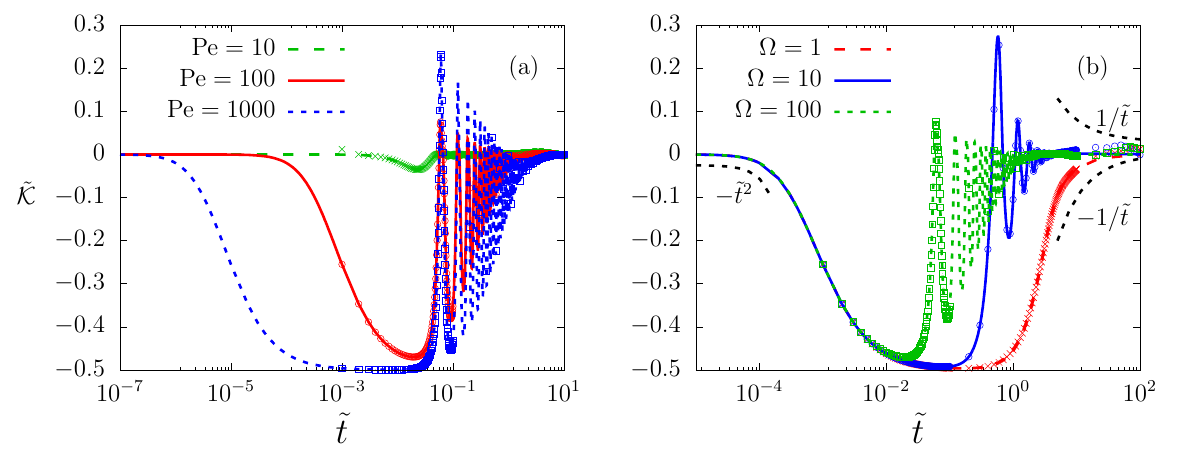}
    \caption{Excess kurtosis of 2d cABPs (a) for different values of $\Pe$ at a fixed $\Omega = 100$ and (b) for different values of $\Omega$ at a fixed $\Pe = 100$. The points are from simulations, and the solid lines denote plots of analytic expressions. The long and short time scaling of $\tilde{\cal{K}}$ are shown in (b) using the dashed lines. As $\tt \to 0$, $\tilde{\cal{K}}$ vanishes as $-\tt^2$. At late times, for $\Omega = 1$, $\tilde{\cal{K}}$ vanishes as $-1/\tt$ while for $\Omega = 10$, $\tilde{\cal{K}}$ vanishes as $1/\tt$. 
    } 
    \label{2d-kurt}
\end{figure}

For the two dimensional case, $\langle  \tilde{r}_i  \tilde{r}_j \rangle = \frac{\delta_{ij}}{2}\langle  \tbr^2\rangle$. The cross-correlation vanishes only if the initial orientation is uniformly distributed (Eq.~\ref{xxyycorr}]), 
reducing the expression of $\tilde{\cal{K}}$ as,
\begin{equation}
    \tilde{\cal{K}} = \frac{\langle{\tbr^4}\rangle}{2 \langle  \tbr^2\rangle^2} - 1
\end{equation}
We plot the excess kurtosis for different values of $\Pe$ and $\Omega$ (Fig.~\ref{2d-kurt}). The plots show the departure from Gaussian at time scales dependent on $\Pe$ and $\Omega$. It is interesting to note that excess kurtosis deviates from zero as time progresses, returning to zero again at a long enough time. At intermediate time scales, it shows oscillations with possibly multiple zero crossings. $\tilde{\cal{K}} \to 0$ at short times is dominated by equilibrium diffusion. At a longer time, active propulsion leads to non-Gaussian departures, and $\tilde{\cal K}$ shows oscillations due to the chiral nature of cABPs. $\tilde{\cal{K}} \to 0$ at $\tt \gg 1$~($t \gg D_r^{-1}$) is due to the decreasing importance of persistence for such long trajectories.
\\
In the short time limit $\tt \to 0$,
\begin{align}
   \lim_{\tt \to 0} \tilde{\cal{K}} =  -\frac{\text{Pe}^4}{32} \tt^2 +\frac{\text{Pe}^4}{192} \left(3 \text{Pe}^2+4\right) \tt^3-\frac{\text{Pe}^4}{23040}\left(135 \text{Pe}^4+360 \text{Pe}^2-120 \Omega ^2+136\right)\tt^4\nonumber 
   +\mathcal{O}\left(\tt^5\right).
\end{align}
It clearly indicates the initial deviation towards negative values of $\tilde{\cal K}$ dominated by $\Pe$ (Fig.~\ref{2d-kurt}). The effect of chirality is seen in the coefficient of $\tt^4$. The $\sim\, -\tt^2$ vanishing of $\tilde{\cal K}$ as $\tt \to 0$ is observable in Fig.~\ref{2d-kurt}(b).\\  
In the long time limit, the excess kurtosis can be obtained as:
\begin{align*}
   \lim_{\tt \to \infty} \tilde{\cal{K}}  = \frac{ \text{Pe}^4 (5\Omega ^2-7)}{\left(\Omega ^2+1\right) \left(\Omega ^2+4\right) \left(\text{Pe}^2+2 \Omega ^2+2\right)^2}\left(\frac{1}{\tt}\right) 
    - \frac{2 \text{Pe}^6 \left(5 \Omega ^4-12 \Omega ^2+7\right)}{\left(\Omega ^2+1\right)^2 \left(\Omega ^2+4\right) \left(\text{Pe}^2+2 \Omega ^2+2\right)^3}\left(\frac{1}{\tt^2}\right) +\mathcal{O}\left(\frac{1}{\tt^3}\right).
\end{align*}
Therefore, as $\tt \to \infty$, $\tilde{\cal{K}} \to 0$ (Fig.~\ref{2d-kurt}). Note that in the long time limit, the vanishing of excess kurtosis could be as $(-1/\tt)$ for $\Omega^2 < 7/5$ or $(1/\tt)$ for $\Omega^2 > 7/5$. This is observed in Fig.~\ref{2d-kurt}(b).

\section{Dynamics of a chiral active Brownian particle in three dimensions (3d)}
Next, we extend our calculations using the Laplace transform method to characterize the dynamics of a cABP in three dimensions, where the chirality is introduced in the form of an external torque acting on an active Brownian particle. 
As before, there is a self-propulsion velocity $v_0$ in the heading direction $\bu$ of the particle and $\bu\cdot\bu = 1$. In the overdamped limit, the equation of motion for the position $\br$ and the heading direction $\bu$ in the presence of an external torque can be written as
\begin{align}
\begin{aligned}
\dot{\textbf{r}} &= v_0\bu+\boldsymbol{\xi}^T(t)
\\
\dot{\bu} &=  [\boldsymbol{\omega}+\boldsymbol{\xi}^R(t)]\times\bu 
\label{langevin}
\end{aligned}
\end{align}
where $\bm{\omega}$ stands for the chirality, $\bm{\xi}^T(t)$ and $\bm{\xi}^R(t)$ are the translational and rotational noise respectively. The noise correlations satisfy:
\begin{align}
\begin{aligned}
\langle \xi^T_{i}(t)\rangle &=0
\\ 
\langle \xi^R_{i}(t)\rangle &=0
\\
\langle\xi^T_{i}(t)\xi^T_{j}(t^\prime)\rangle &=2D\delta_{ij}\delta(t-t^\prime)
\\
\langle\xi^R_{i}(t)\xi^R_{j}(t^\prime)\rangle &=2D_r\delta_{ij}\delta(t-t^\prime).
\end{aligned}
\end{align}

$D$ and $D_r$ are the translational and rotational diffusion coefficients, respectively. The dynamical equation for orientation can be expressed in terms of the spherical angles $\theta$ and $\phi$ (with $\bu= \sin \theta \cos\phi \hat{x}+\sin\theta\sin\phi\hat{y}+\cos\theta\hat{z}$ in the Cartesian representation). The constant angular velocity $\boldsymbol{\omega}$ can be expressed in terms of $(\omega_0, \theta_{\omega},\phi_{\omega})$. In the Cartesian system, $\bm{\omega}= \omega_0\sin \theta_\omega \cos\phi_{\omega}\hat{x}+\omega\sin\theta_\omega\sin\phi_\omega\hat{y}+\omega_0\cos\theta_\omega\hat{z}$ where $\theta_\omega$ and $\phi_\omega$ are constant. A derivation of the Fokker-Planck equation in 3d is shown in Appendix~\ref{app_orientation}. The dynamical equations are
\begin{align}
\begin{aligned}
d\theta(t) &= \omega_0 \sin\theta_\omega \sin(\phi_\omega-\phi)dt+ \frac{D_r}{\tan\theta} dt+\xi _\theta dt
\\
d\phi(t) &=  \omega_0\Big(\cos\theta_\omega-\cot\theta\sin\theta_\omega \cos(\phi_\omega-\phi)\Big)dt+\frac{\xi_\phi dt}{\sin\theta}
\label{theta-phiMain}
\end{aligned}
\end{align}
We set ${\bm{\omega}} = \omega_0 \hat{\bm{z}}$ so that $\theta_\omega=0$; as a result, the above equations reduce to: 
$$ d\theta(t) = \frac{D_r}{\tan\theta} dt+\xi _\theta dt; \,\,\, d\phi(t) =  \omega_0 dt+\frac{\xi_\phi dt}{\sin\theta}.$$


\subsection{Fokker-Planck Equation and Derivation of the Moment Generation Equation}

The Fokker-Planck equation satisfied by the single-particle probability distribution function, $P(\br,\bu,t)$, is given as~(see Appendix~\ref{app_FP_3d} for a detailed derivation):
\begin{align}
\partial_t P &=D\nabla^2 P+D_r\bR^2 P- v_0\bu\cdot\nabla P - {\bm{\omega}}\cdot\bR P 
\label{FPE_main}
\end{align}
where $\bR \equiv \bu\times \bm{\nabla}_{\bu}$ is the rotation operator. 
By applying the Laplace transformation to the Fokker-Planck equation, we get:
\begin{align}
-P(\br,\bu,0)+s\tilde{P}(\br,\bu,s)=D\nabla^2\tilde{P} +D_r\bR^2\tilde{P} - v_0\bu\cdot\nabla\tilde{P} -{\bm{\omega}}\cdot\bR \tilde{P}
\label{LFPE3d}
\end{align}
where $\tilde{P}(\br,\bu,,s)=\int_0^\infty dt e^{-st}P(\br,\bu,t)$ is the Laplace transformation of $P(\br,\bu,t)$ and $P(\br,\bu,0)=\delta^3(\br)\delta(\bu-\bu_0)$ is the initial probability distribution function.
Again, considering an arbitrary dynamic variable which is a function of $\br$ and $\bu$, $\psi=\psi(\br,\bu)$ and multiplying Eq.~\ref{LFPE3d} by $\psi$ and integrating with respect to $\br$ and $\bu$, we have:
\begin{align}
-\langle\psi\rangle_0+s\langle\psi\rangle_s=v_0\langle\bu\cdot\nabla\psi\rangle_s+D\langle\nabla^2\psi\rangle_s+D_r\langle\bR^2\psi\rangle_s+\langle{\bm{\omega}}\cdot\bR\psi\rangle_s
\end{align}
where $\langle\psi\rangle_0=\int d \br\int d \bu P(\br,\bu,0) \psi(\br,\bu)$ and $\langle\psi\rangle_s=\int d \br\int d\bu \tilde{P}(\br,\bu,s) \psi(\br,\bu)$.

For this cyclic swimmer, without loss of generality, we choose the direction of the constant torque along the $z-$ axis, i.e., ${\bm{\omega}} = \omega {\bm{\hat{z}}}$ which leads to a helical trajectory with a circular motion in the $x-y$ plane. The  equation for computing the moments further simplifies to,
\begin{align}
-\langle\psi\rangle_0 +s\langle\psi\rangle_s =v_0\langle\bu\cdot\nabla\psi\rangle_s+D\langle\nabla^2\psi\rangle_s+D_r\langle\bR^2\psi\rangle_s+\omega\langle\bR_z\psi\rangle_s
\label{ME3}
\end{align}
Using the above equation, we can compute the Laplace transformed form of all the moments as shown in the following sections.
 
\numberwithin{equation}{subsection} 
\subsection{Orientation Autocorrelation}
We first choose $\psi=\bu$. Now, from the properties of the operator $\bR$, we can write $\mathcal{R}_\alpha u_\beta=-\epsilon_{\alpha\beta\gamma}u_\gamma$, $\bR^2\psi=-2\psi$ in three dimensions. Furthermore, $\nabla^2 \psi=0 $, $\nabla \psi= 0$ and $\langle \psi \rangle_0=\bu_0$. From Eq.~\ref{ME3} we get,
\begin{align}
(s+2D_r)\langle u_x \rangle_s +\omega \langle u_y \rangle_s &=u_{0x}
\label{u0x3d}
\\
(s+2 D_r)\langle u_y \rangle_s -\omega \langle u_x \rangle_s &=u_{0y}
\label{u0y3d}
\\
(s+2 D_r)\langle u_z \rangle_s  &=u_{0z}
\end{align}
Solving for $\langle u_x\rangle_s$, $\langle u_y\rangle_s$ and $\langle u_z\rangle_s$ and performing the inverse Laplace transform, we get the time evolution of the components of orientation vector, $\bu$ as,
\begin{align}
\begin{aligned}
 \langle u_x (t)\rangle&=e^{-2D_rt}\left[u_{0x}\cos( \omega t)-u_{0y}\sin(\omega t)\right]
\\
\langle u_y (t)\rangle&=e^{-2D_rt}\left[u_{0y}\cos(\omega t)+u_{0x}\sin(\omega t)\right]
\\
\langle u_z (t)\rangle&=u_{0z}e^{-2D_rt}
\end{aligned}
\end{align} 
The time correlation of the orientational vector has the following form
\begin{align}
\langle\bu\cdot\bu_0\rangle=e^{-2D_rt}\left[\cos \omega t+u_{0z}^2(1-\cos \omega t)\right]\, ,
\end{align}
which decays with persistent time $D_r^{-1}$ and has an oscillatory nature for $u_{0z} \neq 1$. 

\numberwithin{equation}{subsection}
\subsection{Mean Displacement}
Considering $\psi=\br$ in Eq.~\ref{ME3}, using the expressions of $\langle u_i\rangle_s$ from the previous section and performing inverse Laplace transform, one can get components of displacement as follows
\begin{align}
\begin{aligned}
&\langle x(t)\rangle=\frac{v_0}{4D_r^2+\omega^2}\Big[a(1-e^{-2D_rt}\cos \omega t)+b e^{-2D_r t}\sin \omega t\Big]
\\
&\langle y(t)\rangle=\frac{v_0}{4D_r^2+\omega^2}\Big[b(1-e^{-2D_rt}\cos \omega t)+ a e^{-2D_r t}\sin \omega t\Big]
\\
&
\langle z(t)\rangle=\frac{v_0 u_{0z}}{2D_r}[1-e^{-2D_rt}]
\end{aligned}
\end{align}
with $a=2D_ru_{0x}-\omega u_{0y}$ and $b=2D_ru_{0y}+\omega u_{0x}$. 


\numberwithin{equation}{subsection}
\subsection{Position-orientation cross-correlation}
Considering $\psi=\br\cdot\bu$ and $\psi=\br \cdot \bu^\perp$ in Eq. \ref{ME3}, we get the following position-orientation equal time cross-correlations :
\begin{align}
    \begin{aligned}
        \langle \br\cdot\bu \rangle &= \frac{v_0}{12}\Big[\frac{\omega ^2 e^{-6 D_r t} \left(1-3 u_{0z}^2\right)}{D_r \left(16 D_r^2+\omega ^2\right)}+\frac{24 D_r^2+2 \omega ^2}{4 D_r^3+D_r \omega ^2}+\frac{3 e^{-2 D_r t} \left(u_{0z}^2-1\right)}{D_r}\Big]
        \\& + \frac{v_0 e^{-2 D_r t} \sin (t \omega ) \left(\omega  \left(-4 D_r^2-\omega ^2\right) u_{0z}^2+\omega  \left(12 D_r^2+\omega ^2\right)\right)}{64 D_r^4+20 D_r^2 \omega ^2+\omega ^4}
        \\& -\frac{v_0 e^{-2 D_r t} \left(4 D_r \left(4 D_r^2+\omega ^2\right) u_{0z}^2 \cos (t \omega )+16 D_r^3 \cos (t \omega )\right)}{64 D_r^4+20 D_r^2 \omega ^2+\omega ^4}
        \\
        \langle \br\cdot\bu^\perp\rangle &= \frac{v_0 e^{-6 D_r t} }{3 \left(64 D_r^4+20 D_r^2 \omega ^2+\omega ^4\right)}\Big(-2 \omega  e^{6 D_r t} \left(16 D_r^2+\omega ^2\right)+\omega  \left(4 D_r^2+\omega ^2\right) \left(3 u_{0z}^2-1\right)\Big)
        \\&+ \frac{v_0 e^{-2 D_r t} \sin (t \omega ) \left(16 D_r^3+4 D_r \left(4 D_r^2+\omega ^2\right) u_{0z}^2\right)}{64 D_r^4+20 D_r^2 \omega ^2+\omega ^4}
        \\&+\frac{v_0 e^{-2 D_r t} \cos (t \omega ) \left(\omega  \left(12 D_r^2+\omega ^2\right)-\omega  \left(4 D_r^2+\omega ^2\right) u_{0z}^2\right)}{64 D_r^4+20 D_r^2 \omega ^2+\omega ^4}
    \end{aligned}
\end{align}
It is important to note that $\bu^\perp= -\bm{\hat{x}} u_y+\bm{\hat{y}} u_x$ is the unit vector perpendicular to both the external torque and the orientation of the instantaneous direction of the particle.

\numberwithin{equation}{subsection}
\subsection{Mean Squared Displacement (MSD)}
\begin{figure}[t]
\centering
  \includegraphics[width=\linewidth]{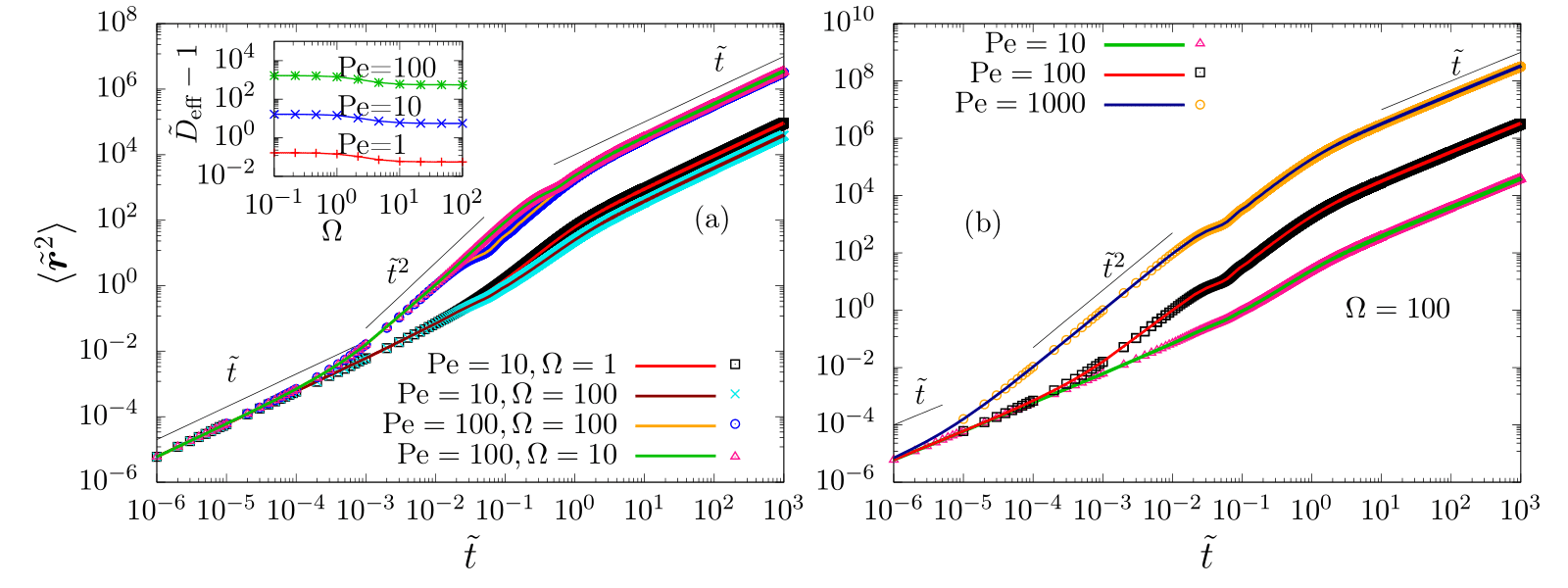}
  \caption{(a) The MSD of 3d cABPs are plotted against time (in the unit of $D_r^{-1}$) for various combinations of $\Omega$ and $\Pe$. The solid lines denote analytical expressions, and the points are obtained from simulations. The plots show oscillations at intermediate times at high $\Omega$. The oscillations are less pronounced than those seen in two dimensions. (Inset) The late-time effective diffusivity shows saturation at different values of $\Pe$ as $\Omega \rightarrow 0$ and decreases with $\Omega$ to saturate to $\Pe^2/18$. (b) MSD for different $\Pe$ at a fixed $\Omega = 100$. With increasing $\Pe$, the initial transition time from diffusive to ballistic crossover $\tt_I$ shifts to lower $\tt$. }
  \label{msdPe3d}
\end{figure}
 \begin{figure}[t]
\centering
  \includegraphics[width=0.5\linewidth]{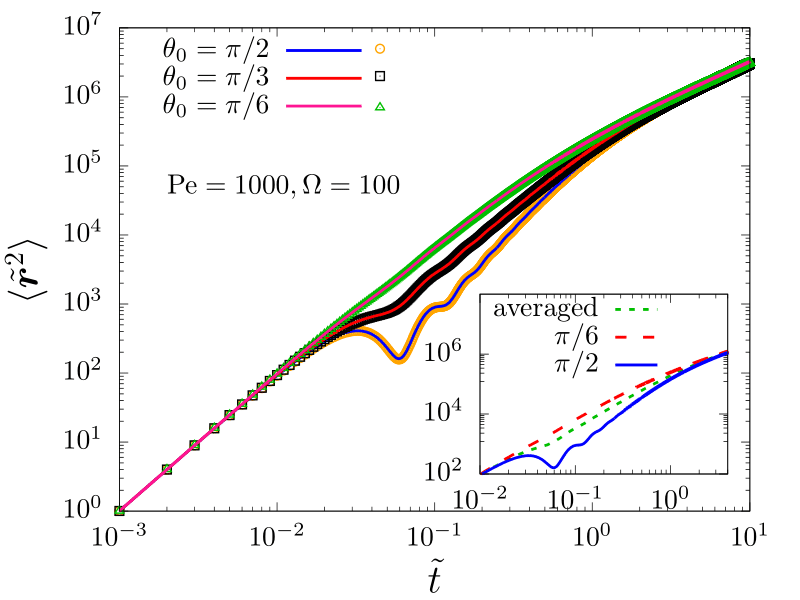}
  \caption{Plot of the MSD with time (in the unit of $D_r^{-1}$) for a given combination of $\Pe$ and $\Omega$ and with varying initial orientation of cABPs. The initial orientation is fixed along a certain direction given by $\theta_0$. Only the component $ z-$ of the initial orientation vector, $u_{0z} = \cos{\theta_0}$, has a profound effect on MSD. In the inset, we compare the MSD when it is averaged over initial orientations with the MSD for fixed initial orientations $\theta_0$. The effect of $\theta_0$ is evident at intermediate time scales. For time scales, $D_r t \gg 1$, the effect vanishes as expected from our analytical results.}
  \label{msd3du0}
\end{figure}
As in two dimensions, we set $\psi = \br^2$ to get the mean square displacement. Eq.~\ref{ME3} gives $s\langle \br^2\rangle_s=6D_t\langle 1\rangle_s+2v_0\langle \br\cdot\bu\rangle_s$. Now,  $\langle 1\rangle_s=1/s$, $\langle \br^2\rangle_s=6D/s^2+2v_0\langle \br\cdot\bu\rangle_s/s$. Putting $\psi = \br\cdot\bu$ in Eq.~\ref{ME3} and using $\nabla\psi=\bu$ and $\bR^2(\br\cdot\bu)=-2\br\cdot\bu$ we get $(s+2D_r)\langle \br\cdot\bu\rangle_s = v_0/s + \omega\langle \bR_z(\br\cdot\bu)\rangle_s$. 

Next, we consider $\psi=\bR_z(\br\cdot\bu)=-(xu_y-yu_x)$. 
\begin{align}
\begin{aligned}
\nabla\bR_z(\br\cdot\bu)&=\bR_z\nabla(\br\cdot\bu)=\bR_z\bu
\\
v_0\bu\cdot\nabla\bR_z(\br\cdot\bu)&=v_0\bu\cdot\bR_z\nabla(\br\cdot\bu)=v_0\hat{\textbf{u}}\cdot\bR_z\bu=0
\\
\bR^2\bR_z(\br\cdot\bu)&=-2\bR_z(\br\cdot\bu)
\\
\bR_z\bR_z(\br\cdot\bu)&=\bR_z(-(xu_y-yu_x))=-(xu_x+yu_y)=-\br\cdot\bu+zu_z
\end{aligned}
\end{align}
Hence from Eq.~\ref{ME3}, we get $(s+2D_r)\langle\bR_z(\br\cdot\bu)\rangle_s=-\omega\langle\br\cdot\bu\rangle_s+\omega \langle zu_z\rangle_s$. This leads to 
\begin{align}
    \langle \br\cdot\bu\rangle_s=\frac{v_0(s+2D_r)}{((s+2D_r)^2+\omega^2)}+\frac{\omega^2 \langle zu_z\rangle_s}{((s+2D_r)^2+\omega^2)}.
\end{align}
Evaluation of $\langle zu_z\rangle_s$ again requires us to set $\psi = \langle zu_z\rangle_s$. We need to evaluate quantities such as $\bR^2u_iu_j$ (which are shown in the Appendix~\ref{app_Rot}) to finally lead to $\langle zu_z\rangle_s=v_0\langle u_z^2\rangle_s/(s+2D_r)$ and $\langle u_z^2\rangle_s=u_{0z}^2/(s+6D_r)+2D_r/(s+6D_r)$. Putting all these expressions back into the equation of $\langle \br^2\rangle_s$ and taking its inverse Laplace transform, we get the following expression for the mean square displacement: 
\begin{align}
\begin{aligned}
\langle \br^2(t)\rangle &= \Big(6D+\frac{v_0^2}{D_r}\frac{D_r^2+\omega^2/12}{D_r^2+\omega^2/4}\Big)t+\frac{v_0^2e^{-2D_rt}}{4D_r^2}(1-u^2_{0z})+\frac{e^{-6D_rt}v_0^2\omega^2(3u_{0z}^2-1)}{36D_r^2(16D_r^2+\omega^2)}
\\
&+\frac{v_0^2\Big(4D_r^2(3u_{0z}^2-1)\omega^2+\omega^4(3u_{0z}^2-4)-144D_r^4\Big)}{18D_r^2(4D_r^2+\omega^2)^2}
\\
&-e^{-2D_rt}\sin\omega t\frac{4D_r\omega v_0^2\Big(4D_r^2(5+u_{0z}^2)+\omega^2(1+u_{0z}^2)\Big)}{(4D_r^2+\omega^2)^2(16D_r^2+\omega^2)}
\\
&+e^{-2D_rt}\cos \omega t\frac{2v_0^2\Big(32D_r^4(u_{0z}^2+1)+(12D_r^2\omega^2+\omega^4)(u_{0z}^2-1)\Big)}{(4D_r^2+\omega^2)^2(16D_r^2+\omega^2)}
\end{aligned}
\label{rsq3dt}
\end{align}


It is evident from the above expression that $\langle \br^2 \rangle$ depends on the initial orientation of the particle, $u_{0z}$. If we average over the initial orientations of the particle, which is equivalent to replacing $u_{0z}^2$ by the mean $\langle u_{0z}^2 \rangle = 1/3$, the MSD reduces to
\begin{align}
\begin{aligned}
\langle \br^2(t)\rangle &= \left[6D+\frac{v_0^2}{D_r}\frac{D_r+\omega^2/12}{D_r+\omega^2/4}\right]t +
\frac{v_0^2}{6D_r^2}e^{-2D_rt} - \frac{v_0^2}{6D_r^2(4D_r^2+\omega^2)^2}\left[ \omega^4 + 48D_r^4\right]
\\
&-\frac{16D_r\omega v_0^2}{3(4D_r^2+\omega^2)^2} e^{-2D_rt}\sin\omega t 
+\frac{4v_0^2(4D_r^2 - \omega^2)}{3(4D_r^2+\omega^2)^2}e^{-2D_rt}\cos \omega t
\end{aligned}
\label{rsq3davg}
\end{align}
which is the same as obtained in Ref.~\cite{sevilla2016diffusion}. In the asymptotic limit of $D_r t \gg 1$ and $ \omega t \gg 1$, the MSD shows diffusive behavior, $\langle \br^2 \rangle(t) = 6 D_{\rm eff} t$ with the effective diffusion coefficient,
\begin{align}
    D_{\rm eff}=D+\frac{v_0^2}{6D_r}\frac{1+\Omega^2/12}{1+\Omega^2/4}.
\end{align}
In the limit of a vanishing external torque $\Omega \rightarrow 0$, the above expression reduces to the well-known behavior of simple ABPs. Using $\Pe$, the equation can be expressed as
\begin{align}
    \tilde{D}_{\rm eff}=\frac{D_{\rm eff}}{D} = 1 + \frac{\Pe^2}{6}\frac{1+\Omega^2/12}{1+\Omega^2/4}
\end{align}
which is the same as obtained in Ref.~\cite{sevilla2016diffusion}.
This is represented in Fig.~\ref{msdPe3d}(a, Inset). To analyze crossovers at short times, we expand the MSD as $t\to 0$ to give:
\begin{align}
\lim_{t \to 0} \langle\br^2\rangle &= 6 D t+t^2 v_0^2-\frac{2}{3} t^3 \left(v_0^2 D_r\right)+\frac{1}{12} t^4 v_0^2 \left(4 D_r^2+\left(u_{0z}^2-1\right) \omega ^2\right)+\mathcal{O}\left(t^5\right).
\end{align}
In the short time regime, the MSD exhibits diffusive dynamics, $\langle\br^2\rangle = 6D t$ which crosses over to ballistic motion, $\langle\br^2\rangle \simeq v_0^2 t^2$, at $t_{I}\approx 6D/v_0^2$, as shown in Figure~\ref{msdPe3d}. The effect of chirality in MSD appears in the coefficient of the term $t^4$, suggesting that the effect manifests itself in dynamics over relatively longer periods. The dependence of the MSD on the initial orientation can be observed in Fig.~\ref{msd3du0} at intermediate time scales. When averaged over all possible initial orientations, we get: 
\begin{align}
    \begin{aligned}
        \langle \tilde \br^2(\tilde t) &= \left(6 + \frac{\text{Pe}^2 \left(\Omega ^2+12\right)}{3 \left(\Omega ^2+4\right)}\right)\tilde t  
        \\
        &+ \frac{\text{Pe}^2 e^{-2 \tt} \left(-e^{2 \tilde t} \left(\Omega ^4+48\right)-8 \left(\Omega ^2-4\right) \cos (\Omega \tt )-32 \Omega  \sin (\Omega \tt )+\left(\Omega ^2+4\right)^2\right)}{6 \left(\Omega ^2+4\right)^2}.
    \end{aligned}
\end{align}
This is also shown in Fig.~\ref{msd3du0}(inset). 

%

\subsubsection{MSD along and perpendicular to the external torque}
Considering $\psi=z^2$, we can compute the MSD along the direction of the external torque:
\begin{align}
    \begin{aligned}
     \left\langle \br_{\parallel}^2(t)\right\rangle=   \left\langle z^2(t)\right\rangle &= \frac{e^{-6 t D_r} \left(3 v_0^2 u_{0z}^2-v_0^2\right)}{36 D_r^2}+\frac{e^{-2 t D_r} \left(v_0^2-v_0^2 u_{0z}^2\right)}{4 D_r^2}\\
     & +\frac{t \left(6 D_r D+v_0^2\right)}{3 D_r}+\frac{3 v_0^2 u_{0z}^2-4 v_0^2}{18 D_r^2} 
    \end{aligned}
\end{align}
which gives the asymptotic diffusive behavior $\left\langle \br_{\parallel}^2(t)\right\rangle = 2 D_\parallel t$ with
\begin{align}
    \frac{D_\parallel}{D} = 1 + \frac{\Pe^2}{6}.
\end{align}

The MSD in the perpendicular $x-y$ plane has an oscillatory nature, which can be computed by subtracting the expression of $\langle z^2 \rangle$ from the expression of $\langle \br^2\rangle$:
\begin{align}
    \begin{aligned}
        \langle \br^2_{\perp}(t)\rangle & = \left[4 D + \frac{8 v_0^2 D_r}{3 \left(4 D_r^2+\omega ^2\right)}\right]t
        \\&-\frac{4 v_0^2 \left(3 u_{0z}^2-1\right) e^{-6 t D_r}}{9 \left(16 D_r^2+\omega ^2\right)}- \frac{2 v_0^2 \left(4 D_r^2 \left(3 u_{0z}^2+5\right)+\omega ^2 \left(3 u_{0z}^2-7\right)\right)}{9 \left(4 D_r^2+\omega ^2\right)^2}
        \\& +\cos(\omega t)\frac{2 v_0^2 e^{-2 t D_r} \left(12 \omega ^2 D_r^2 \left(u_{0z}^2-1\right)+32 D_r^4 \left(u_{0z}^2+1\right)+\omega ^4 \left(u_{0z}^2-1\right)\right)}{\left(4 D_r^2+\omega ^2\right){}^2 \left(16 D_r^2+\omega ^2\right)}
        \\&+\sin(\omega t) \frac{2 v_0^2 e^{-2 t D_r} \left(-2 \omega ^3 D_r \left(u_{0z}^2+1\right)-8 \omega  D_r^3 \left(u_{0z}^2+5\right)\right)}{\left(4 D_r^2+\omega ^2\right){}^2 \left(16 D_r^2+\omega ^2\right)}
    \end{aligned}
\end{align}
In the long-time limit, the approximate expression of $ \langle \br^2_{\perp}(t)\rangle = 4 D_{\perp} t$ with~\cite{sevilla2016diffusion} 
\begin{align}
    \tilde{D}_\perp = \frac{D_{\perp}}{D}
=1+\frac{\Pe^2}{6}\frac{4}{4+\Omega^2}.
\end{align}
\begin{figure}[t]
\centering
  \includegraphics[width=\linewidth]{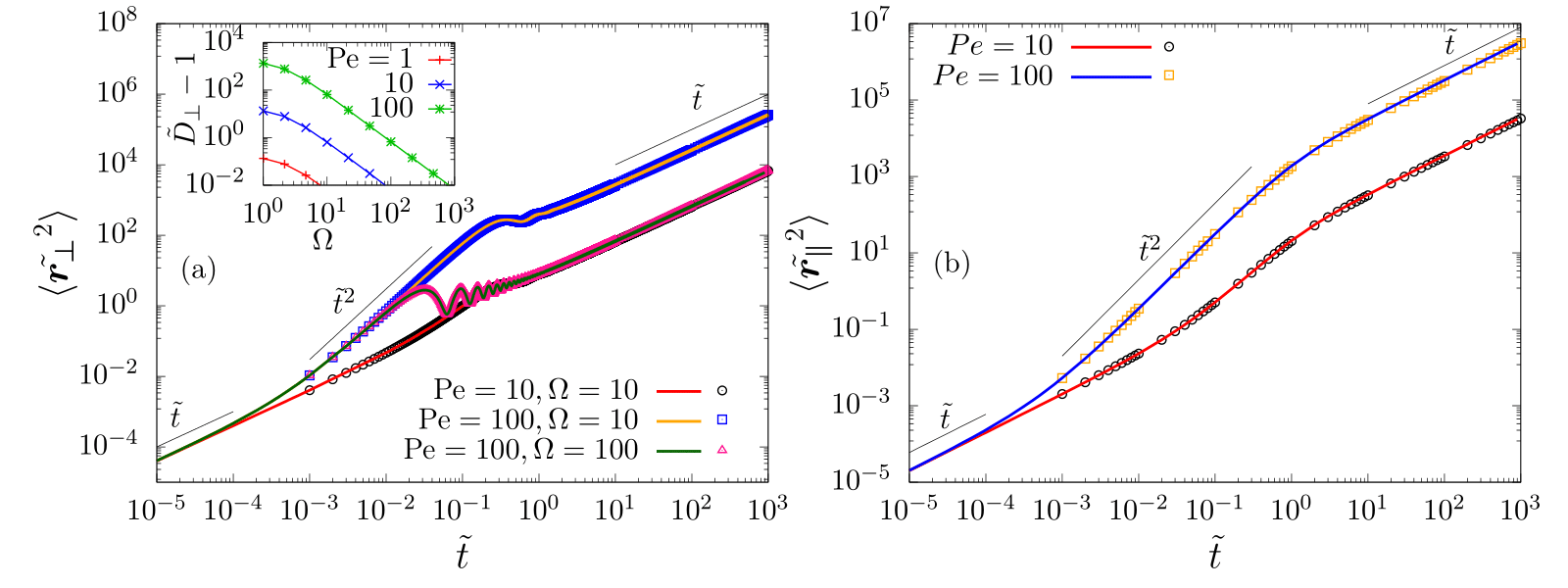}
  \caption{(a) MSD in the plane perpendicular to the direction of torque for various $\Pe$ and $\Omega$. (Inset) Effective diffusivity $\tilde{D}_\perp$ as a function of $\Omega$ for three different values of $\Pe$ show $1/\Omega^2$ scaling at large $\Omega$. (b) MSD with time in the torque direction for two different values of $\Pe$ at $\Omega = 100$. Diffusive-ballistic crossovers are shown in both. In (a) and (b), solid lines denote analytical plots, and the points are from simulations.}
  \label{msdPerp3d}
\end{figure}
This is plotted in Fig.~\ref{msdPerp3d}(a)(inset) as a function of $\Omega$ for various values of $\Pe$. 
It is interesting to note that the diffusion coefficient in the $x-y$ plane is not the same as $D_{\rm eff}$ in two dimensions. Moving from two to three dimensions impacts diffusivity $D_\perp$ in a non-trivial way due to the persistence of cABPs. The amount of chirality has a significant impact on their dynamics. In the limit of large $\Omega$, the active part of the diffusivity gets vanishingly small, and the total diffusivity approaches the equilibrium value.



We can again express $\langle \tilde\br_{\parallel}^2\rangle$ and $\langle \tilde\br_{\perp}^2\rangle$ in dimensionless form:
\begin{align}
    \begin{aligned}
       \left\langle \tilde\br_{\parallel}^2(\tilde t)\right\rangle= & \tilde t\left(\frac {6 + \Pe^2}{3}\right) + \frac{\text{Pe}^2}{36}  e^{-6 \tilde t} \left(-9 e^{4\tilde t} \left(u_{0z}^2-1\right)+e^{6 \tilde t} \left(6 u_{0z}^2-8\right)+3 u_{0z}^2-1\right)
       \\
       \left\langle \tilde\br_{\perp}^2(\tilde t)\right\rangle= & \tilde t\left(\frac{8 \text{Pe}^2}{3 \left(\Omega ^2+4\right)}+4\right)+ \frac{2 \text{Pe}^2 \left(3 u_{0z}^2 \left(\Omega ^2+4\right)-7 \Omega ^2+20\right)}{9 \left(\Omega ^2+4\right)^2} + \frac{4 \text{Pe}^2 \left(1-3 u_{0z}^2\right)e^{-6\tilde t }}{9 \left(\Omega ^2+16\right)}
       \\& + \frac{2 \text{Pe}^2  \left(u_{0z}^2 \left(\Omega ^2+4\right) \left(\Omega ^2+8\right)-\Omega ^2 \left(\Omega ^2+12\right)+32\right)e^{-2\tilde t}\cos{\Omega\tilde t }}{\left(\Omega ^2+4\right)^2 \left(\Omega ^2+16\right)}
       \\&  -\frac{4 \text{Pe}^2  \Omega  \left(u_{0z}^2 \left(\Omega ^2+4\right)+\Omega ^2+20\right)e^{-2\tilde t}\sin{\Omega t}}{\left(\Omega ^2+4\right)^2 \left(\Omega ^2+16\right)}.
    \end{aligned}
\end{align}
In Fig.~\ref{msdPerp3d}, we plot this dimensionless MSD along and perpendicular to the torque for different values of $\Pe$ and $\Omega$. The results show multiple diffusive-ballistic crossovers and an exact match with the simulations.



\numberwithin{equation}{subsection}
\subsection{Fourth Moment}
To evaluate the kurtosis, we require the expression of the fourth moment of the displacement. To evaluate the fourth moment we put $\psi = \br^4$ in Eq.~\ref{ME3} to get:
\begin{align}
s\left\langle \br^{4}\right\rangle_s=4(d+2) D\left\langle \br^{2}\right\rangle_s+4 v_{0}\left\langle \br^{2} \br\cdot\bu \right\rangle_{s}
\label{r4moment3d}
\end{align}
Expression of $\left\langle \br^{2}\right\rangle_s$ is obtained from the previous section. The second term can be calculated following the steps given below. 

\subsubsection{Calculation of $\left\langle \br^{2} \br\cdot\bu\right\rangle_{s}$}
Choosing $\psi=\br^{2} \br\cdot\bu$ in Eq.~\ref{ME3}, we get:
\begin{align}
\begin{aligned}
\left[s+2D_r+\frac{\omega^{2}}{s+2D_r}\right]\langle \br^2 \br\cdot \bu\rangle_s = 10D\left[\langle\br \cdot \bu\rangle_{s}+\frac{\omega\langle \bR_{z}(\br\cdot\bu)\rangle_{s}}{s+2D_r}\right]
\\ + v_{0}\left[2 \left\langle(\br\cdot\bu)^{2}\right\rangle_{s} 
+\frac{\omega\langle \bR_{z}(\br\cdot\bu)^{2}\rangle_{s}}{s+2D_r}+\langle \br^{2}\rangle_{s}\right]
+\frac{\omega^{2}\langle \br^{2} z u_{z}\rangle_{s}}{s+2D_r}
\end{aligned}
\end{align}

Choosing  $\psi = \br^{2} zu_z$ in Eq.~\ref{ME3}, we get:
\begin{align}
\begin{aligned}
\left[s+2D_{r}\right]\left\langle \br^{2}zu_{z}\right\rangle_{s}&=  10D\left\langle z u_{z}\right\rangle_{s}+2 v_{0}\left\langle\br\cdot\bu zu_{z}\right\rangle_{s}+v_{0}\left\langle \br^{2} u_{z}^{2}\right\rangle_{s}
\end{aligned}
\end{align}
For $\left\langle\br\cdot\bu zu_{z}\right\rangle_{s}$,
\begin{align}
\nonumber
\left[s+6 D_{r}+\frac{\omega^{2}}{s+6D_{r}}\right]\left\langle\br\cdot\bu z u_{z}\right\rangle_{s} = 2 D\left\langle u_{z}^{2}\right\rangle_{s}+2 D_{r}\left\langle z^{2}\right\rangle_{s}+v_{0}\left\langle z u_{z}\right\rangle_{s}+\frac{\omega^{2}\left\langle z^{2} u_{z}^{2}\right\rangle_{s}}{s+6D_{r}}\\
+v_{0}\left[\left\langle\br\cdot\bu u_{z}^{2}\right\rangle_{s}+\frac{\omega\left\langle \bR_{z}\left(\br\cdot\bu u_{z}^{2})\right\rangle_{s}\right.}{s+6D_{r}}\right]
\end{align}
with $\langle \bR_{z}\left(\br\cdot\bu z u_{z}\right)\rangle_{s}$ evaluated as:
\begin{align}
\begin{aligned}
{\left(s+6 D_{r}\right)\left\langle \bR_{z}\left(\br\cdot\bu z u_{z}\right)\right\rangle_{s}=v_{0}\left\langle \bR_{z}\left(\br\cdot \bu u_{z}^{2}\right)\right\rangle_{s} } -\omega\left\langle\br\cdot \bu z u_{z}\right\rangle_{s}+\omega\left\langle z^{2} u_{z}^{2}\right\rangle_{s}.
\end{aligned}
\end{align}
For $\left\langle\br \cdot \bu u_{z}^{2}\right\rangle_{s}$,
\begin{align}
\begin{aligned}
\left[s+12D_{r}+\frac{\omega^{2}}{s+12D_{r}}\right]\left\langle\br\cdot\bu u_{z}^{2}\right\rangle_{s} = 2 D_{r}\langle\br\cdot\bu\rangle_{s}+4 D_{r}\left\langle z u_{z}\right\rangle_{s}+v_{0}\left\langle u_{z}^{2}\right\rangle_{s} \\
+\frac{\omega^{2}\left\langle z u_{z}^{3}\right\rangle_{s}}{s+12D_{r}}+\frac{2 D_{r}\omega\left\langle \bR_{z}(\br\cdot \bu)\right\rangle_{s}}{s+12D_{r}}
\end{aligned}
\end{align}
with $\langle \bR_{z}\left(\br\cdot\bu\right)\rangle_{s}$ evaluated as:
\begin{align}
\begin{aligned}
{\left[s+12D_{r}\right]\left\langle R_{z}\left(\br\cdot \bu u_{z}^{2}\right)\right\rangle_{s} =  2D_{r}\left\langle R_{z}(\br\cdot \bu)\right\rangle_{s} } &-\omega\left\langle(\br\cdot \bu) u_{z}^{2}\right\rangle_{s}
+\omega\left\langle  z u_{z}^{3}\right\rangle_{s}
\end{aligned}
\end{align}
With $\psi = zu_z^3$ in Eq.~\ref{ME3}, we get:
\begin{align}
\left[s+12D_{r}\right]\left\langle z u_{z}^{3}\right\rangle_{s}=6 D_{r}\left\langle z u_{z}\right\rangle_{s}+v_{0}\left\langle u_{z}^{4}\right\rangle_{s}
\end{align}
and similarly
\begin{align}
    \left[s+20D_{r}\right]\left\langle u_{z}^{4}\right\rangle_{s}=12\left\langle u_{z}^{2}\right\rangle_{s}+u_{z0}^4
\end{align}
\begin{align}
\left(s+6D_{r}\right)\left\langle z^{2} u_{z}^{2}\right\rangle_{s}=2 D\left\langle u_{z}^{2}\right\rangle_{s}+2 D_{r}\left\langle z^{2}\right\rangle_{s}+2 v_{0}\left\langle z u_{z}^{3}\right\rangle_{s}
\end{align}
\begin{align}
\left(s+6D_{r}\right)\left\langle \br^{2} u_{z}^{2}\right\rangle_{s}=2 D_{r}\left\langle \br^{2}\right\rangle_{s}+2 d D\left\langle u_{z}^{2}\right\rangle_{s}+2 v_{0}\left\langle\br . \bu u_{z}^{2}\right\rangle_{s}
\end{align}
\begin{figure}[t]
\centering
  \includegraphics[width=\linewidth]{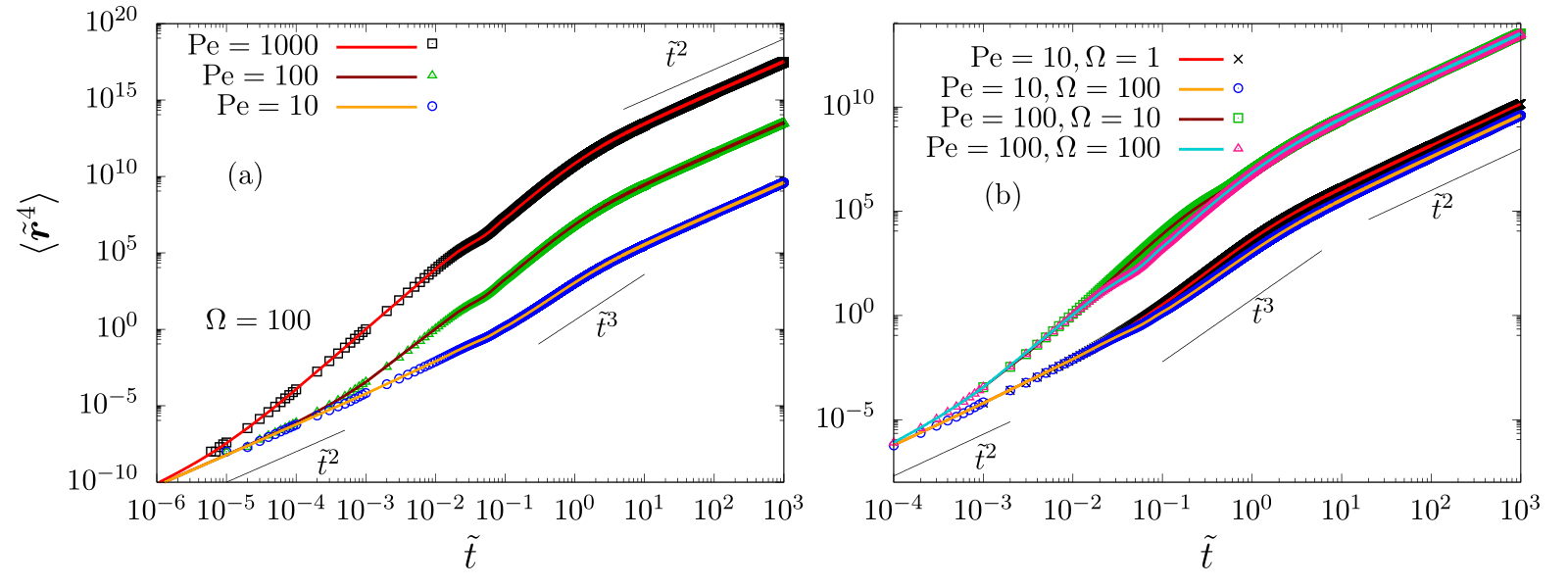}
  \caption{The fourth moment of displacements plotted against time (in the unit of $D_r^{-1}$) for (a) various $\Pe$ at a fixed $\Omega = 100$ and (b) various combinations of $\Pe$ and $\Omega$. Multiple crossovers are observed from $~t^2$ to $~t^3$. The $\tt \to \infty$ limit depends on both $\Omega$ and $\Pe$. In both (a) and (b), the solid lines denote analytical formulae, and the points are obtained from simulations.}
  \label{r4Pe3d}
\end{figure}
Now choosing $\psi = \br\cdot\bu$, we get:
\begin{align}
\begin{aligned}
\left[s+6D_{r}+\frac{4 \omega^{2}}{s+6D_r}\right]\left\langle(\br\cdot\bu)^{2}\right\rangle_{s}=\frac{2 D}{s}+2 D_{r}\left\langle \br^{2}\right\rangle_{s}+2 v_{0}\langle\br\cdot\bu\rangle_{s} +\frac{2 \omega v_{0}\left\langle \bR_{z}(\br\cdot\bu)\right\rangle_{s}}{s+6D_{r}} 
\\ 
+ \frac{6 \omega^{2}\left\langle\br\cdot\bu z u_{z}\right\rangle_{s}}{s+6D_{r}}+ \frac{2 \omega^{2}\left\langle \br^{2}\right\rangle_{s}}{s+6D_{r}} - \frac{2 \omega^{2}\left\langle z^{2}\right\rangle_{s}}{s+6D_{r}}-\frac{2 \omega^{2}\left\langle \br^{2} u_{z}^{2}\right\rangle_{s}}{s+6D_{r}}
\end{aligned}
\end{align}
and $\langle \bR_{z}(\br\cdot\bu)^{2}\rangle_{s}$ can be obtained from:
\begin{align}
\begin{aligned} {\left(s+6D_{r}\right) \left\langle \bR_{z}(\br\cdot\bu)^{2}\right\rangle_{s} } &=2 v_{0}\left\langle \bR_{z}(\br\cdot\bu)\right\rangle_{s}-4 \omega\left\langle(\br\cdot\bu)^{2}\right\rangle_{z} +6 \omega\left\langle\br\cdot\bu z u_{z}\right\rangle _{s}\\ &+2 \omega\left\langle \br^{2}\right\rangle_{s}-2 \omega\left\langle z^{2}\right\rangle_{s} -2 \omega\left\langle \br^{2} u_{z}^{2}\right\rangle_{s} \end{aligned}
\end{align}
Using all these expressions, one can calculate $\langle \br^4(t)\rangle$ taking the inverse Laplace transform of Eq.~\ref{r4moment3d}. This expression depends on the initial orientation. Considering the initial orientation as uniformly distributed and averaging over all such orientations, we obtain the dimensionless expression of the fourth moment in terms of $\text{Pe}$ and $\Omega$ as shown in Eq.\eqref{r43dtavg} of Appendix~\ref{app_r4_3d}. In the absence of chirality, $\Omega = 0$, this reduces to 
\begin{align}
    \begin{aligned}
       \langle \tbr^4(t)\rangle &=  \frac{5}{3} \left(\Pe^2+6\right)^2 \, \tt^2+\frac{1}{9} \Pe^2 \left(-9 \left(\Pe^2-10\right) e^{-2 \tt}-26 \Pe^2-90\right) \tt
       \\&+\frac{1}{54} \Pe^4 e^{-6 \tt} \left(-108 e^{4 \tt}+107 e^{6 \tt}+1\right), \nonumber
    \end{aligned}
\end{align}
a result that agrees with earlier derivations~\cite{shee2020active}.
%

The fourth moment of cABP is plotted for various values of $\Pe$ and $\Omega$ in Fig.~\ref{r4Pe3d} along with the simulation results. 
In the short-time limit, expanding the above expression around $\tt = 0$, we get:
\begin{align}
     \lim_{\tt \to 0} \langle\tbr^4\rangle & = 60 \tt^2+20 \Pe^2 \tt^3+\left(\Pe^4-\frac{40 \Pe^2}{3}\right) \tt^4-\frac{2}{9} \Pe^2 \tt^5 \left(6 \Pe^2 + 5 \Omega ^2 -30\right)\nonumber\\ &+ \mathcal{O}\left(\tt^6\right).
\end{align}
As in the two-dimensional case, the fourth moment exhibits $\langle\tbr^4\rangle \simeq 60\tt^2$, dominated by thermal fluctuations, which then crosses over to $\langle\tbr^4\rangle\simeq 20 \text{Pe}^2 \tt^3$, dominated by high activity, at $\tt_{I}=3/\text{Pe}^2$~(Fig.~\ref{r4Pe3d}). Note that this crossover is independent of $\Omega$ and is common to cABP and ABP. The chiral rotation manifests itself only in the coefficient of the further higher-order term $\tt^5$, suggesting that chirality influences the dynamics only at relatively longer time regimes. 

The late-time behavior $\tt\to\infty$ of the fourth moment can be obtained from the following expansion
\begin{align}
     \lim_{\tt \to \infty} \langle\tbr^4\rangle & = \frac{1}{9} \left[\frac{\Pe^4 \left(3 \Omega ^4 + 40 \Omega ^2 + 240\right)}{\left(\Omega ^2 + 4\right)^2}+\Pe^2 \left(\frac{480}{\Omega ^2+4}+60\right)+540\right] \tt^2
\end{align}
showing the $\langle\tbr^4\rangle\sim \tt^2$ scaling as observed in Fig.~\ref{r4Pe3d}, where the coefficient depends on both activity $\text{Pe}$ and chirality $\Omega$.

\begin{figure}
    \centering
    \includegraphics[width=\linewidth]{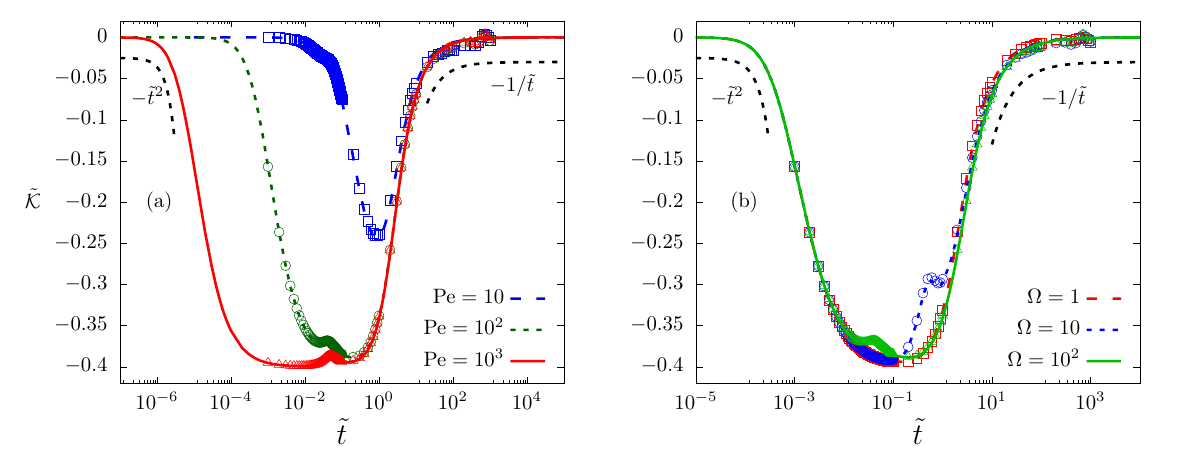}
    \caption{Excess kurtosis for 3d cABPs in the presence of an external torque (a) for different values of $\Pe$ at a fixed $\Omega = 100$ and (b) for different values of $\Omega$ at a fixed $\Pe = 100 $. The exact analytical results (lines) are compared with the simulation results (points). The dependence on the time scales near $\tt \to 0$ and $\tt \to \infty$ is also shown in the figure.}
    \label{kurt-3d}
\end{figure}

\subsection{Excess Kurtosis:  Deviation from Gaussian }
We can calculate the excess kurtosis of cABP in three dimensions. In this case, Eq.~\ref{mu4_avg} becomes $2\langle  r_i  r_j \rangle^2=2\Big( \langle  x^2 \rangle^2 +\langle  y^2 \rangle^2+ \langle  z^2 \rangle^2 +\langle  x y \rangle^2+\langle  yz \rangle^2+\langle  z x \rangle^2\Big) $.
The expression of $\langle z^2\rangle$ is easily calculated.  Averaging the initial orientations, we can conclude that all cross terms vanish. The calculation of all the displacement correlation functions is shown in Appendix~\ref{3d_2ndorder_corr}. 

We can calculate the excess kurtosis for various values of $\Pe$ and $\Omega$ as shown in Fig.~\ref{kurt-3d}. The analytical expressions agree perfectly with the results of the numerical simulation. At sufficiently short times, $\tilde{\cal K}$ remains vanishingly small, dominated by thermal fluctuations. It then deviates from zero to negative values as a result of the active non-Gaussian departures. At long enough times, it vanishes again as the persistence time becomes negligibly small compared to the time elapsed, $\tt \gg 1$. On intermediate time scales, oscillations can be observed because of the chiral rotations of cABPs. However, unlike in two dimensions, the trajectories in three dimensions extend in a direction perpendicular to the direction of chiral rotation, thus suppressing such oscillations.
    
In the short time limit, we expand the excess kurtosis around $\tt=0$ to obtain the form,
\begin{align}
 \begin{aligned}
    \tilde {\cal{K}}&=-\frac{\text{Pe}^4 }{90}\tt^2+\frac{1}{270} \text{Pe}^4 \left(\text{Pe}^2+4\right) \tt^3 \\& -\frac{\left(\text{Pe}^4 \left(15 \left(\text{Pe}^2+8\right) \text{Pe}^2-20 \Omega ^2+152\right)\right)}{16200}\tt^4 + O\left(\tt^5\right)\, .
 \end{aligned}
\end{align}
Thus $\tilde {\cal{K}}$ vanishes in the limit $\tt \to 0$ as $\tilde {\cal{K}} \sim - \tt^2$~(see Fig.~\ref{kurt-3d}).  The initial deviations from Gaussianity towards negative kurtosis values are controlled by $\Pe^4$. The effect of chirality is seen only later in terms of the coefficient of the $\tt^4$ term in the above expansion. 

In the long time limit, the approximate expression of excess kurtosis can be obtained using an expansion around $\tt=\infty$, 
\begin{align}
    \begin{aligned}
      \begin{split}
        \tilde{\cal{K}} &= -\left(\frac{1}{\tt}\right)\left[\frac{\text{Pe}^4 }{ 15  \left(\Omega ^2+4\right) \left(\Omega ^2+9\right) \left(\Omega ^2+36\right) } \times \right.
        \\& \left. \frac{\left(33 \Omega ^{10}+1841 \Omega ^8+30344 \Omega ^6+228192 \Omega ^4+777600 \Omega ^2+3421440\right)}{ \left(40 \left(\text{Pe}^2+6\right) \left(\text{Pe}^2+18\right) \Omega ^2+240 \left(\text{Pe}^2+6\right)^2+3 \left(\text{Pe}^4+20 \text{Pe}^2+180\right) \Omega ^4\right)}\right]
        \\&
        -\left(\frac{1}{\tt^2}\right)\left[\frac{\text{Pe}^6 \left(33 \Omega ^{10}+1841 \Omega ^8+30344 \Omega ^6+228192 \Omega ^4+777600 \Omega ^2+3421440\right)}{15 \left(\Omega ^2+4\right)^2 \left(\Omega ^2+9\right) \left(\Omega ^2+36\right)}\times \right.
        \\& \left.
        \frac{\text{Pe}^2 \left(\Omega ^2+12\right) \left(3 \Omega ^4+80\right)+30 \left(\Omega ^2+4\right) \left(\Omega ^4+48\right)}{\left(40 \left(\text{Pe}^2+6\right) \left(\text{Pe}^2+18\right) \Omega ^2+240 \left(\text{Pe}^2+6\right)^2+3 \left(\text{Pe}^4+20 \text{Pe}^2+180\right) \Omega ^4\right)^2}\right]\\& + O\left(\frac{1}{\tt^3}\right).
     \end{split}
    \end{aligned}
\end{align}
This expression clearly shows that in the limit of $\tt \to \infty$, the excess kurtosis vanishes as $-1/\tt$. 

\section{Discussion}
In this study, we extensively examined the behavior of free chiral active Brownian particles (cABP) subject to thermal noise. Our approach involved utilizing the Laplace transform of the Fokker–Planck equation, enabling us to derive expressions for the time-dependent dynamical moments of various observables in both two- and three-dimensions. We derived exact forms for different displacement moments and equal-time cross-correlations with orientations. This allowed us to determine the impacts of persistence, chirality, and dimensionality on the dynamics. For example, while the MSD and fourth moments of displacement in 2d turn out to be independent of initial orientation, the initial orientation, along with the chirality, influences these quantities nontrivially in three dimensions. Using series expansions, we further illustrated the various time-dependent scaling behavior of the dynamical variables and their crossovers at various time regimes. To confirm the accuracy of our analytical predictions, we performed rigorous validation using direct numerical simulations.  

In addition, we utilized analytic calculations of excess kurtosis to determine the deviations from the Gaussian behavior of the displacement variable. The disappearance of excess kurtosis in the short- and long-time limits is extracted from series expansions around $\tt=0$ and $\infty$. We found a scaling of $(-\tt^2)$ as $\tt \to 0$ in two- and three-dimensions. In the long time limit, the nature of the vanishing of excess kurtosis depends on the dimension. In two dimensions, it can go to zero as $(1/\tt)$ or $(-1/\tt)$, depending on the magnitude of the chiral rotation. In contrast, in three dimensions, it always shows a $(-1/\tt)$ scaling at long $\tt$.  

This detailed consideration of the nature of excess kurtosis for a cABP adds further intrigue when considering the difference in the behavior of a cABP versus that of a chiral active Ornsetein-Uhlenbeck particle (cAOUP)~\cite{caprini2019active, Caprini2023}.  The phenomenology of cABPs and cAOUPs is similar, and the results are the same at the second moment. They differ in the higher moments, and a non-zero excess kurtosis in cABP differentiates the two. Since cAOUPs follow the Gaussian process, their excess kurtosis is necessarily zero. 
It is, therefore, imperative to experimentally measure the excess kurtosis of cABPs to determine which model is best suited to describe the observed spatiotemporal behavior.

\section*{Acknowledgments}
AP and AC thanks the computing facility at IISER Mohali. DC thanks Abhishek Dhar for collaborations on related topics, acknowledges a research grant from the Department of Atomic Energy (1603/2/2020/IoP/R\&D-II/150288), and thanks ICTS-TIFR, Bangalore, for an Associateship. AS thanks Cristián Huepe for general discussions on the effects of chirality in active systems and acknowledges partial support from the John Templeton Foundation, Grant 62213.

\appendixpage

\begin{appendices}
\renewcommand{\thesection}{\Roman{section}}
\renewcommand{\thesubsection}{\thesection.\Roman{subsection}}
\renewcommand{\thesubsubsection}{\thesubsection.\Roman{subsubsection}}
\section{Derivation of the Fokker-Planck Equation in two dimensions (2d)}\numberwithin{equation}{section}
\label{app_FP_2d}
\noindent
In Ito formalism, the equation of motion of a chiral active Brownian particle in two dimensions (Eq.~\ref{langevin2d}) can be alternatively expressed as,
\begin{align}
    \begin{aligned}
        d{\br} &= v_0\bu dt+d \bm{B}
\\
d{\phi} &= \omega dt+ d W
    \end{aligned}
\end{align}
$\bm{B}$ and $W$ are translational Wiener processes and they have zero mean and correlations, $\langle dB_{i}dB_{j}\rangle=2D \delta_{ij}dt$, $\langle dW^2 \rangle=2D_r dt$ and $\bm{B}$ and $W$ are independent of each other. The Taylor series expansion of an arbitrary function of $\br$ and $\phi$, $F[\br(t),\bu(t)]$, 
\begin{align}
    \begin{aligned}
         dF=\nabla_i F  d r_i+\partial_\phi F  d \phi+ \frac{1}{2}\left(\nabla_i\nabla_j F d r_i dr_j+\nabla_i\partial_\phi F  d r_i d \phi+\partial_\phi^2 F  d \phi^2\right)+...
          \label{SDE1_2d}
    \end{aligned}
\end{align}
The average of Eq.~\ref{SDE1_2d} satisfies,
\begin{align}
\begin{aligned}
    \langle dF\rangle &= \left\langle v_0 u_i\nabla_i F+\omega\frac{\partial F}{\partial \phi} +D\nabla^2 F+D_r\partial^2_\phi F \right\rangle dt
    \\
   \Rightarrow \frac{\langle dF\rangle}{dt}& = \left\langle v_0 u_i\nabla_i F+\omega\frac{\partial F}{\partial \phi} +D\nabla^2 F+D_r\partial^2_\phi F \right\rangle
    \label{SDE2_2d}
\end{aligned}
\end{align}
All terms higher order than $dW^2$, $dB^2$ and $dt$ are neglected. $\frac{\langle dF\rangle}{dt}$ can be alternatively expressed as, $\frac{\langle dF\rangle}{dt}=\left \langle \frac{dF}{dt}\right\rangle=\frac{d}{dt} \int   d \br  d\phi F[\br(t),\bu(t)] P[(\br,\phi,t\vert\br_0,\phi_0,0)]=\int   d \br  d\phi \frac{dP}{dt} F$, where $P[(\br,\phi,t\vert\br_0,\phi_0,0)]$ is the conditional probability density function of $\br$ and $\phi$. Therefore, Eq~\ref{SDE2_2d} takes the form,
\begin{align}
\begin{aligned}
 \int   d \br  d\phi F\partial_t P &=  v_0 \int   d \br  d\phi P \bu\cdot \nabla F +\omega \int   d \br  d\phi P\partial_\phi F+ D  \int   d \br  d\phi P \nabla^2 F
 \\&+ D_r \int   d \br  d\phi \partial^2_\phi F P
 \end{aligned}
\end{align}
Integrating by parts and discarding all the surface terms,
\begin{align}
\begin{aligned}
 \int   d \br  d\phi F\partial_t P &=  -v_0 \int   d \br  d\phi F \bu\cdot \nabla P -\omega \int   d \br  d\phi F\partial_\phi P+ D  \int   d \br  d\phi F \nabla^2 P
 \\&+ D_r \int   d \br  d\phi \frac{\partial^2 P}{\partial^2 \phi} F
 \end{aligned}
\end{align}
For an arbitrary $F$, the probability density function satisfies the following Fokker-Planck equation,
\begin{align}
    \partial_t P&= D \nabla^2 P + D_r \partial^2_\phi P - v_0 \bu\cdot\nabla P-\omega \partial_\phi P 
\end{align}
which is the same as Eq.~\ref{FPE}. 

\section{Derivation of the Fokker-Planck Equation in three dimensions (3d)}\numberwithin{equation}{section}
\label{app_FP_3d}
\noindent
In Ito formalism, the equation of motion of an ABP under an external torque (Eq.~\ref{langevin}) can be expressed as,
\begin{align}
    \begin{aligned}
        d \br & = v_0 \bu dt + d \bm{B}
        \\
        d \bu &= \bm{\omega}\times \bu dt+ d\bm{W}\times \bu
        \label{Ito1}
    \end{aligned}
\end{align}
  The translational and rotational Wiener processes satisfy $\langle dB_{i}dB_{j}\rangle=2D \delta_{ij}dt$, $\langle dW_{i}dW_{j}\rangle=2D_r \delta_{ij}dt$ and $\langle dB_{i}\rangle=0=\langle dW_{i}\rangle$. $\bm{B}$ and $\bm{W}$ do not correlate with each other. Consider an arbitrary function $f[\br(t),\bu(t)]$. Taylor series expansion of $f$, 
  \begin{align}
          df=\frac{\partial f}{\partial r_i}  d r_i+\frac{\partial f}{\partial u_i}  d u_i+ \frac{1}{2}\left(\frac{\partial^2 f}{\partial r_i\partial r_j}  d r_i dr_j+\frac{\partial^2 f}{\partial r_i\partial u_j}  d r_i d u_j+\frac{\partial^2 f}{\partial u_i\partial u_j}  d u_i du_j\right)+...
          \label{SDE1}
  \end{align}
 Incorporating Eq.~\ref{Ito1} and using the properties of the Wiener processes we compute the average of eq.\ref{SDE1},
  \begin{align}
  \nonumber
          \langle df \rangle&=\Big\langle \frac{\partial f}{\partial r_i}  ( v_0 u_i dt + d B_i)+\frac{\partial f}{\partial u_i}  \epsilon_{imn}(\omega_m dt+ d W_n) u_n
       \\ \nonumber&+ \frac{1}{2}\left(\frac{\partial^2 f}{\partial r_i\partial r_j}  ( v_0 u_i dt + d B_i) ( v_0 u_j dt + d B_j)+\frac{\partial^2 f}{\partial r_i\partial u_j}  ( v_0 u_i dt + d B_i) \epsilon_{jkl}(\omega_k dt+ d W_k) u_l\right.
       \\ \nonumber& \left.+\frac{\partial^2 f}{\partial u_i\partial u_j}  \epsilon_{imn}(\omega_m dt+ d W_m) u_n \epsilon_{jkl}(\omega_k dt+ d W_k)\right)\Big\rangle
  \\
  \Rightarrow
  \frac{\langle df\rangle}{dt}&=\left\langle\frac{\partial f}{\partial r_i}  v_0 u_i +\epsilon_{imn}\omega_m u_n\frac{\partial f}{\partial u_i}+ D  \frac{\partial^2 f}{\partial r^2} +D_r \epsilon_{mni}\epsilon_{mlj}u_l u_n \frac{\partial^2 f}{\partial u_i\partial u_j}\right\rangle
          \label{SDE2}
  \end{align}
  All terms of order higher than $dt$, $dB^2$, and $dW^2$ are neglected. Let us consider the rotational operator, $\bR\coloneqq \bu\times\frac{\partial}{\partial \bu}$. The second and fourth terms can be written in terms of $\bR$:
  \begin{align}
      \epsilon_{imn}\omega_m u_n\frac{\partial f}{\partial u_i} &= \omega_m  \epsilon_{mni}u_n\frac{\partial f}{\partial u_i} =\bm{\omega}\cdot \bR f
      \\
      \nonumber
      \epsilon_{mni}\epsilon_{mlj}u_l u_n \frac{\partial^2 f}{\partial u_i\partial u_j}&= \epsilon_{mlj}u_l \bR_m \left(\frac{\partial f}{\partial u_j}\right) =\epsilon_{mlj}\bR_m \left(u_l  \frac{\partial f}{\partial u_j}\right) - \epsilon_{mlj}\left(\bR_m u_l \right) \frac{\partial f}{\partial u_j}
      \\ \nonumber& = \bR_m \epsilon_{mlj}\left(u_l  \frac{\partial f}{\partial u_j}\right)+ \epsilon_{mlj}\epsilon_{mlk} u_k \frac{\partial f}{\partial u_j} 
      \\ \nonumber & = \bR_m  \bR_m f+2\delta_{jk}u_k \frac{\partial f}{\partial u_j}
      \\ \nonumber &=\bR^2 f-2\bu \cdot \frac{\partial f}{\partial \bu} 
      \\&=\bR^2 f
  \end{align}
 Here, we have used the property of the unit vector $\bu \cdot \frac{\partial f}{\partial \bu} =0$ and that of the rotational operator $\bR_\alpha u_\beta=-\epsilon_{\alpha\beta\gamma} u_{\gamma}$~\cite{doi1988theory}. We now obtain the simplified equation satisfied by $\langle f\rangle$:
  \begin{align}
      \frac{\left\langle df\right\rangle}{dt} =v_0 \left\langle\bu\cdot\nabla f   \right\rangle+ \left\langle \bm{\omega}\cdot \bR f\right\rangle+ D\langle \nabla ^2 f\rangle+D_r \left\langle \bR^2 f \right\rangle
     \label{SDE4}
  \end{align}
 $\langle f \rangle=\int   d \br  d\bu f[\br(t),\bu(t)] P[(\br,\bu,t\vert\br_0,\bu_0,0)]$ where $P$ is the conditional probability density function. Now,
  \begin{align}
  \begin{aligned}
   & \frac{d \langle f \rangle}{dt} =\left\langle \frac{df}{dt}\right\rangle =  \frac{d}{dt} \int   d \br  d\bu f[\br(t),\bu(t)] P[(\br,\bu,t\vert\br_0,\bu_0,0)]=\int   d \br  d\bu f\frac{dP}{dt}
    \\
   & v_0 \left\langle\bu\cdot\nabla f\right\rangle   = -v_0 \int   d \br  d\bu f \bu\cdot\nabla P 
    \\
   & \left\langle \bm{\omega}\cdot \bR f\right\rangle = -\int   d \br  d\bu f  \bm{\omega}\cdot \bR P 
    \\
   & D\langle \nabla ^2 f\rangle   =D \int   d \br  d\bu f \nabla^2 P 
   \\
   &
   D_r \left\langle \bR^2 f \right\rangle =D_r \int   d \br  d\bu  f\bR^2 P 
   \label{elements}
   \end{aligned}
  \end{align}
  Substituting all the expressions of Eq \ref{elements} into Eq. \ref{SDE4} we can obtain the Fokker-Planck equation satisfied by the probability density function $P$,
  \begin{align}
      \partial_t P &=D\nabla^2 P+D_r \bR^2 P- v_0\hat{\textbf{u}}\cdot\nabla P-\bm{\omega}\cdot\bR P
  \end{align}
which is the same as Eq.~\ref{FPE_main}.
  
\section{Evolution of angles of orientation vector}\numberwithin{equation}{section}
\label{app_orientation}
\noindent
 The orientation vector $\hat{u}$ satisfies following overdamped Langevin equation,
\begin{align}
 \begin{aligned}
\dot{\bu}&=[\bm{\omega}+\boldsymbol{\xi}^R]\times\bu
\label{Langevin_u}
\end{aligned}
\end{align}
The noise terms satisfy
\begin{align}
\begin{aligned} 
\langle \xi_{Ri}(t)\rangle &=0
\\
\langle\xi_{Ri}(t)\xi_{Tj}(t^\prime)\rangle &=2D_r\delta_{ij}\delta(t-t^\prime)
\end{aligned}
\end{align}
Equation \ref{Langevin_u} can be considered as a Brownian motion performed by a unit vector on the surface of a unit sphere. So $\bu$ can be expressed in terms of the angles $\theta$ and $\phi$ in spherical co-ordinates, $\bu= \sin \theta \cos\phi \hat{x}+\sin\theta\sin\phi\hat{y}+\cos\theta\hat{z}$. Hence one can write spherical angles, $\theta$ and $\phi$ in terms of $u_x$, $u_y$ and $u_z$.
\begin{align}
\begin{aligned}
\theta &= \tan^{-1}\left(\frac{\sqrt{u_x^2+u_y^2}}{u_z}\right)
\\
\phi &= \tan^{-1}\left(\frac{u_y}{u_x}\right)
\label{theta-phi}
\end{aligned}
\end{align} 
The evolution of an arbitrary function, $g[\textbf{x}(t)]$ of a stochastic process $\textbf{x}(t)$ can be written as \cite{gardiner1985handbook},
\begin{align}
dg[\textbf{x}(t)]=\frac{\partial g}{\partial t}dt+ \frac{\partial g}{\partial x_i}dx_i+\frac{1}{2}\frac{\partial^2 g}{\partial x_i \partial x_j}dx_i dx_j+...
\end{align}
The stochastic variable, $\textbf{x}$ satisfies the differential equation, $d\textbf{x}=
a(t)+b(t)d\textbf{W}$ with $dW_i=\eta_i dt$ a Wiener process. We neglect terms higher order than $dt$ and $dW_i^2$.

Now, the orientation vector of a chiral active Brownian particle can be described by,
 \begin{align}
 \begin{aligned}
d u_i&=\epsilon_{ijk} (\omega_j dt+dW_j) u_k
\end{aligned}
\end{align}
The Wiener process satisfies $dW_idW_j=2D_r \delta_{ij}dt$ and $dW_i=0$. The constant angular velocity $\boldsymbol{\omega}$ can be expressed as $(\omega_0 \sin\theta_{\omega}\cos\phi_{\omega},\omega_0 \sin\theta_{\omega}\sin\phi_{\omega},\omega_0 \cos\theta_{\omega})$
Hence, $\theta$ and $\phi$ can be expressed as the following stochastic differential equations.
\begin{align}
d\theta(t) &= \omega_0 \sin\theta_\omega \sin(\phi_\omega-\phi)dt+ \frac{D_r}{\tan\theta} dt+dW_\theta
\\
d\phi(t) &=  \omega_0\Big(\cos\theta_\omega-\cot\theta\sin\theta_\omega \cos(\phi_\omega-\phi)\Big)dt+\frac{dW_\phi}{\sin\theta}
\end{align}
$dW_\theta$ and $dW_\phi$ are independent Wiener processes that satisfy $dW^2_\theta=2 D_r dt=dW^2_\phi$. They can be achieved by following transformations,
\begin{align}
\begin{aligned}
dW_\theta&= \cos\phi dW_y-\sin\phi dW_x
\\
dW_\phi &=\sin\theta dW_z-\cos\theta(\cos\phi dW_x+\sin\phi dW_y)
\end{aligned}
\end{align}
\section{Derivation of all the second-order moments}\label{3d_2ndorder_corr}
\subsection{Calculation of $\left\langle z ^2\right\rangle_s$}
\begin{align}
   & s \left\langle z ^2\right\rangle_s =2v_0 \langle zu_z\rangle_s+ 2 D/s
   \\
   & \Big(s+2 D_r \Big) \langle zu_z\rangle_s= v_0 \langle u_z^2\rangle_s
   \\
   & \Big(s+6 D_r \Big) \langle u_z^2\rangle_s= u_{0z}^2+ 2 D_r /s
\end{align}
The final expression $\left\langle z ^2\right\rangle_s=\frac{2 D}{s^2}+\frac{2 v_0^2 \left(\frac{2 D_r}{s}+u_{0z}^2\right)}{s(2 D_r+s) (6 D_r+s)}$.  Averaging over all possible initial orientations and taking inverse Laplace transform,
\begin{align}
    \langle z^2\rangle(t)= \frac{v_0^2 e^{-2 t D_r}}{6 D_r^2}+t \left(\frac{v_0^2}{3 D_r}+2 D\right)-\frac{v_0^2}{6 D_r^2}
\end{align}
\subsection{Calculation of $\left\langle x ^2\right\rangle_s$}
\begin{align}
   & s \left\langle x ^2\right\rangle_s =2v_0 \langle x u_x\rangle_s+ 2 D/s
   \\
   \nonumber
   & \Big(s+2 D_r \Big) \langle xu_x\rangle_s= v_0 \langle u_x^2\rangle_s+\omega \langle \mathcal{R}_z x u_x \rangle_s=v_0 \langle u_x^2\rangle_s-\omega \langle  x u_y \rangle_s
   \end{align}
   This leads to
   \begin{equation}
     \Big(s+2 D_r \Big) \langle xu_x\rangle_s +\omega \langle  x u_y \rangle_s= v_0 \langle u_x^2\rangle_s
   \label{xux}  
   \end{equation}
Similarly $\psi= x u_y$ gives,
\begin{equation}
    \Big(s+2 D_r \Big) \langle xu_y\rangle_s -\omega \langle  x u_x \rangle_s= v_0 \langle u_x u_y\rangle_s
\label{xuy}
\end{equation}
Solving Eq. [\ref{xux}] and [\ref{xuy}] simultaneously,
\begin{align}
    \langle  x u_x \rangle_s\Big[\Big(s+2 D_r \Big)^2+\omega ^2\Big]&= v_0 \Big[\Big(s+2 D_r \Big)\langle u_x^2\rangle_s- \omega \langle u_x u_y\rangle_s\Big]
    \\
    \langle  x u_y \rangle_s\Big[\Big(s+2D_r\Big)^2+\omega ^2\Big]&= v_0 \Big[\Big(s+2 D_r \Big)\langle u_x u_y\rangle_s+ \omega \langle u_x^2\rangle_s\Big]
\end{align}
Now, for $\langle u_x^2\rangle_s$ and $\langle u_x u_y\rangle_s$,
\begin{align}
\nonumber
   & \Big(s+6 D_r \Big) \langle u_x^2\rangle_s= u_{0x}^2+ 2 D_r /s- 2\omega \langle u_x u_y\rangle_s
   \\
   \Rightarrow & \Big(s+6 D_r \Big) \langle u_x^2\rangle_s+2\omega \langle u_x u_y\rangle_s=u_{0x}^2+ 2 D_r /s 
   \\
   \nonumber & \textrm{Considering $\psi= u_x u_y$},
   \\ \nonumber
   & \Big(s+6 D_r \Big) \langle u_x u_y\rangle_s= u_{0x}u_{0y}+\omega \langle u_x^2\rangle_s -\omega \langle u_y^2\rangle_s=u_{0x}u_{0y}+2 \omega \langle u_x^2\rangle_s-\omega/s +\omega \langle u_z^2\rangle_s
   \\
   \Rightarrow &  \Big(s+6 D_r \Big) \langle u_x u_y\rangle_s- 2 \omega \langle u_x^2\rangle_s=u_{0x}u_{0y}-\omega/s +\omega \langle u_z^2\rangle_s
   \\ \nonumber
   &\textrm{The expressions of $\langle u_x^2\rangle_s$ and $\langle u_x u_y\rangle_s$,}
   \\
   & \Big[\Big(s+6 D_r \Big)^2+4 \omega ^2\Big] \langle u_x^2\rangle_s = \Big(s+6 D_r \Big)\Big(u_{0x}^2+ 2 D_r /s\Big)-2 \omega\Big(u_{0x}u_{0y}-\omega/s +\omega \langle u_z^2\rangle_s\Big)
   \\&
   \Big[\Big(s+6 D_r \Big)^2+4 \omega ^2\Big] \langle u_x u_y\rangle_s = \Big(s+6 D_r \Big)\Big(u_{0x}u_{0y}-\omega/s +\omega \langle u_z^2\rangle_s\Big)+2 \omega\Big(u_{0x}^2+ 2 D_r /s\Big)
\end{align}
Here we have used $u_x^2+u_y^2+u_z^2=1$ and in Laplace space, $\langle u_y^2\rangle_s =1/s-\langle u_x^2\rangle_s-\langle u_z^2\rangle_s$
The final expression of $\langle x^2\rangle_s$,
\begin{align}
\begin{aligned}
&  \langle x^2\rangle_s= \frac{2 D}{s^2} + \frac{2 v_0^2 \left((2 D_r+s) (6 D_r+s) \left(2 D_r+s u_{0x}^2\right)-s \omega  u_{0x} u_{0y} (10 D_r+3 s)\right)}{s^2 \left((2 D_r+s)^2+\omega ^2\right) \left((6 D_r+s)^2+4 \omega ^2\right)}
\\
&
+\frac{2 \omega ^2 v_0^2 \left(16 D_r^2-2 D_r s \left(6 u_{0x}^2+5 u_{0z}^2-9\right)+s^2 \left(-2 u_{0x}^2-3 u_{0z}^2+3\right)\right)}{s^2 (6 D_r+s) \left((2 D_r+s)^2+\omega ^2\right) \left((6 D_r+s)^2+4 \omega ^2\right)}
\end{aligned}
\end{align}
Averaging over initial orientations,
\begin{align}
    \langle x^2\rangle_s= \frac{2 \left((2 D_r+s) \left(6 D_r D+3 D s+v_0^2\right)+3 D \omega ^2\right)}{3 s^2 \left((2 D_r+s)^2+\omega ^2\right)}
\end{align}
Taking inverse Laplace transform,
\begin{align}
\begin{aligned}
    \langle x^2\rangle(t) &= \frac{2 v_0^2 e^{-2 t D_r} \left(e^{2 t D_r} \left(\omega ^2-4 D_r^2\right)-\left(\omega ^2-4 D_r^2\right) \cos (t \omega )-4 \omega  D_r \sin (t \omega )\right)}{3 \left(4 D_r^2+\omega ^2\right){}^2}
    \\ & + t\left(\frac{4 v_0^2 D_r}{3 \left(4 D_r^2+\omega ^2\right)}+2 D\right)
    \end{aligned}
\end{align}
\subsection{Calculation of $\left\langle y ^2\right\rangle_s$} 
Using $r^2=x^2+y^2+z^2$, we can write $\left\langle y ^2\right\rangle_s=\left\langle r ^2\right\rangle_s-\left\langle x ^2\right\rangle_s-\left\langle z ^2\right\rangle_s $
\begin{align}
    \begin{aligned}
      \langle y^2\rangle_s &=  \frac{2 D}{s^2}+ \frac{2 v_0^2 \left(s \omega  u_{0x} u_{0y} (10 D_r+3 s)+(2 D_r+s) (6 D_r+s) \left(2 D_r+s u_{0y}^2\right)\right)}{s^2 \left((2 D_r+s)^2+\omega ^2\right) \left((6 D_r+s)^2+4 \omega ^2\right)}
      \\
      & +\frac{2 \omega ^2 v_0^2 \left(16 D_r^2-2 D_r s \left(6 u_{0y}^2+5 u_{0z}^2-9\right)+s^2 \left(-2 u_{0y}^2-3 u_{0z}^2+3\right)\right)}{s^2 (6 D_r+s) \left((2 D_r+s)^2+\omega ^2\right) \left((6 D_r+s)^2+4 \omega ^2\right)}
    \end{aligned}
\end{align}
Taking average over initial orientations,
\begin{align}
   \langle y^2\rangle_s= \frac{2 \left((2 D_r+s) \left(6 D_r D+3 D s+v_0^2\right)+3 D \omega ^2\right)}{3 s^2 \left((2 D_r+s)^2+\omega ^2\right)}
\end{align}
Taking inverse Laplace transform,
\begin{align}
    \begin{aligned}
        \langle y^2\rangle(t) &= \frac{2 v_0^2 e^{-2 t D_r} \left(e^{2 t D_r} \left(\omega ^2-4 D_r^2\right)-\left(\omega ^2-4 D_r^2\right) \cos (t \omega )-4 \omega  D_r \sin (t \omega )\right)}{3 \left(4 D_r^2+\omega ^2\right){}^2}
        \\
        &+ t\left(\frac{4 v_0^2 D_r}{3 \left(4 D_r^2+\omega ^2\right)}+2 D\right)
    \end{aligned}
\end{align}
\subsection{Calculation of $\left\langle xy\right\rangle_s$} 
\begin{align}
    &s \langle xy\rangle_s = v_0 \Big(\langle x u_y\rangle_s +\langle yu_x\rangle_s\Big)
\end{align}
The expression of $\langle x u_y\rangle_s$ is already found in the previous section. $y u_x$ satisfies the following pair of equations,
\begin{align}
    &\Big(s+2 D_r \Big) \langle yu_x\rangle_s +\omega \langle  y u_y \rangle_s= v_0 \langle u_x u_y\rangle_s
    \\
    &
    \Big(s+2 D_r\Big) \langle yu_y\rangle_s -\omega \langle  yu_x \rangle_s= v_0 \langle  u_y^2\rangle_s 
    \\
    \nonumber
    & \textrm{ $\langle yu_x\rangle_s$ satisfies,}
    \\
    &\langle  y u_x \rangle_s\Big[\Big(s+2 D_r \Big)^2+\omega ^2\Big]= v_0 \Big[\Big(s+2D_r\Big)\langle u_x u_y\rangle_s- \omega \langle u_y ^2\rangle_s\Big]
\end{align}
The final expression of  $\langle xy\rangle_s$,
\begin{align}
    \langle xy\rangle_s= \frac{v_0^2 \left(\omega  (10 D_r+3 s) (u_{0x}^2-u_{0y}^2) +2 u_{0x} u_{0y} (2 D_r+s) (6 D_r+s)-4 \omega ^2 u_{0x} u_{0y}\right)}{s \left((2 D_r+s)^2+\omega ^2\right) \left((6 D_r+s)^2+4 \omega ^2\right)}
\end{align}
Averaging over all possible initial conditions, we get $\langle xy\rangle_s=0$.

\subsection{Calculation of $\left\langle xz\right\rangle_s$} 
\begin{align*}
    &s \langle xz\rangle_s = v_0 \Big(\langle x u_z\rangle_s +\langle zu_x\rangle_s\Big)
    \\
    &
    \Big(s+2 D_r\Big) \langle zu_x\rangle_s +\omega \langle  z u_y \rangle_s= v_0 \langle u_x u_z\rangle_s
    \\
    &
    \Big(s+ 2 D_r \Big) \langle zu_y\rangle_s -\omega \langle  zu_x \rangle_s= v_0 \langle  u_y u_z\rangle_s 
    \end{align*} 
    Solving the above equations for $\langle  z u_x \rangle_s$ and $\langle  z u_y \rangle_s$,
    \begin{align*}
    &\langle  z u_x \rangle_s\Big[\Big(s+ 2 D_r \Big)^2+\omega ^2\Big]= v_0 \Big[\Big(s+ 2 D_r Big)\langle u_x u_z\rangle_s- \omega \langle u_z u_y \rangle_s\Big]
     \\
    &\langle  z u_y \rangle_s\Big[\Big(s+2 D_r \Big)^2+\omega ^2\Big]= v_0 \Big[\Big(s+2 D_r\Big)\langle u_y u_z\rangle_s+ \omega \langle u_z u_x \rangle_s\Big]
    \\
    &\Big(s+6 D_r \Big) \langle u_x u_z\rangle_s+  \omega \langle u_z u_y\rangle_s=u_{0x}u_{0z}
    \\
    &\Big(s+6 D_r \Big) \langle u_y u_z\rangle_s-  \omega \langle u_z u_y\rangle_s=u_{0y}u_{0z}
\end{align*}
If we average over all possible initial orientations, $\langle u_{0x}u_{0z}\rangle=0=\langle u_{0y}u_{0z}\rangle$, which gives $\langle u_x u_z\rangle_s=0=\langle u_y u_z\rangle_s$ resulting in $\langle  z u_x \rangle_s=0=\langle  z u_y \rangle_s$.

$\langle x u_z\rangle_s$ satisfies,
\begin{align}
    \Big(s+2 D_r \Big) \langle x u_z\rangle_s = v_0 \langle u_x u_z\rangle_s=0
\end{align}
Hence, $\langle xz\rangle_s=0$ if averaged over initial conditions.

\subsection{Calculation of $\left\langle yz\right\rangle_s$} 
\begin{align}
    &s \langle yz\rangle_s = v_0 \Big(\langle y u_z\rangle_s +\langle z u_y\rangle_s\Big)
\end{align}
Again $\langle y u_z\rangle_s$ satisfies,
\begin{align}
    \Big(s+2 D_r \Big) \langle y u_z\rangle_s = v_0 \langle u_y u_z\rangle_s=0
\end{align}
In the previous section, we derived the expression of $\langle z u_y \rangle_s$. 
$\langle yz\rangle_s=0$ if averaged over initial conditions.
\section{Some important expressions}\numberwithin{equation}{section}
\label{app_Rot}
\noindent
Here are some important expressions, we encountered in the course of the calculation of the fourth moment of the displacement, $\langle \br^4\rangle$. 
\subsection{$\bR^2 u_i u_j u_k$}
We first note that $\bR^{2}\left(u_{i} u_{j} u_{k}\right) 
= \bR_{\alpha} \bR_{\alpha} u_{i} u_{j} u_{k}$. Now,
\begin{align}
\begin{aligned}
& \bR_{\alpha}\left(u_{i} u_{i} u_{k}\right) 
=\left(\bR_{\alpha} u_{i}\right) u_{i} u_{k}+u_{i}\left(\bR_{\alpha} u_{i}\right) u_{k}+u_{i} u_{j} \bR_{\alpha} u_{k} \\
=&-\epsilon_{\alpha i l} u_l u_{j} u_{k}-\epsilon_{\alpha j m} u_{i} u_{m} u_{k}-\epsilon_{\alpha k n} u_{i} u_{j} u_{n}
\end{aligned}
\end{align}
Therefore, to evaluate $\bR^2 u_i u_j u_k$, we need to calculate the following terms as shown below:
First term:
\begin{align}
\begin{aligned}
&\text{First term,}
\\&-\bR_{\alpha}\left(\epsilon_{\alpha i l} u_{l} u_{j} u_{k}\right)\\
&=\epsilon_{\alpha i l} \epsilon_{\alpha l o} u_{o} u_{j} u_{k}+\epsilon_{\alpha i l} \epsilon_{\alpha j p} u_{l} u_{p} u_{k}+\epsilon_{\alpha i l} \epsilon_{\alpha k q} u_{L} u_{j} u_{q}\\
&=\left(\delta_{il} \delta_{l o}-\delta_{i o} \delta_{l l} \right)u_{o} u_{j} u_{k}+\left(\delta_{i j} \delta_{l p}-\delta_{i p} \delta_{l j}\right) u_{l} u_{p} u_{k}+\left(\delta_{i x} \delta_{k q}-\delta_{i q} \delta_{k k}\right) u_{i} u_{j} u_{q}\\
&=-(d-1) u_{i} u_{j} u_{k}+\delta_{i j} u_{k}-u_{i} u_{j} u_{k}+\delta_{i k} u_{i}-u_{i} u_{j} u_{k}\\
&=-(d+1) u_i u_{j} u_{k}+\delta_{i j} u_{k}+\delta_{k i} u_{j}
\\
& \text{Second term,}
\\&-\bR_{\alpha}\left(\epsilon_{\alpha j m} u_{i} u_{m} u_{k}\right)\\
&= \epsilon_{\alpha j m} \epsilon_{\alpha i l} u_{l} u_{m} u_{k}+\epsilon_{\alpha j m} \epsilon_{\alpha m o} u_{i} u_{o} u_{k}+\epsilon_{\alpha j m} \epsilon_{\alpha k p} u_{i} u_{m} u_{p}\\
&=\left(\delta_{i j} \delta_{l m}-\delta_{j l} \delta_{i m}\right) u_{l} u_{m} u_{k}+\left(\delta_{j m} \delta_{m o}-\delta_{j o} \delta_{m m}\right) u_{i} u_{o} u_{k}+\left(\delta_{j k} \delta_{m o}-\delta_{j p} \delta_{m k}\right) u_{i} u_{m} u_{p} u_{j o}\\
&=\delta_{i j}  u_{k}-u_{i} u_{j} u_{k}-(d-1) u_{i} u_{j} u_{k}+\delta_{j k} u_{i}-u_{i} u_{i} u_{k}
\\
&=-(d+1) u_i u_{j} u_{k}+\delta_{i j} u_{k}+\delta_{j k} u_{i}
\\
&-\bR_{\alpha}\left(\epsilon_{\alpha j m} u_{i} u_{m} u_{k}\right)\\
&= \epsilon_{\alpha j m} \epsilon_{\alpha i l} u_{l} u_{m} u_{k}+\epsilon_{\alpha j m} \epsilon_{\alpha m o} u_{i} u_{o} u_{k}+\epsilon_{\alpha j m} \epsilon_{\alpha k p} u_{i} u_{m} u_{p}\\
&=\left(\delta_{i j} \delta_{l m}-\delta_{j l} \delta_{i m}\right) u_{l} u_{m} u_{k}+\left(\delta_{j m} \delta_{m o}-\delta_{j o} \delta_{m m}\right) u_{i} u_{o} u_{k}+\left(\delta_{j k} \delta_{m o}-\delta_{j p} \delta_{m k}\right) u_{i} u_{m} u_{p} u_{j o}\\
&=\delta_{i j}  u_{k}-u_{i} u_{j} u_{k}-(d-1) u_{i} u_{j} u_{k}+\delta_{j k} u_{i}-u_{i} u_{i} u_{k}
\\&= -(d+1) u_i u_{j} u_{k}+\delta_{i j} u_{k}+\delta_{j k} u_{i}
\\
&\text{Third term,}
\\
&\bR_{\alpha}(\epsilon_{\alpha k n} u_{i} u_{j} u_{n})=-(d+1) u_i u_{j} u_{k}+\delta_{k i} u_{j}+\delta_{j k} u_{i}
\end{aligned}
\end{align} 
where $d=3$ is the dimensionality of the embedding space. 
Finally, adding them,
\begin{align}
\bR^2 u_i u_j u_k=-3(d+1) u_i u_{j} u_{k}+2\left(\delta_{k i} u_{j}+\delta_{j k} u_{i}+\delta_{i j} u_{k}\right).
\end{align}
\subsection{$\bR^2 u_a u_b u_c u_d$}
Using the same procedure as before,
\begin{align}
\bR^{2} u_{a} u_{b} u_{c} u_{d}
=-4(d+2) u_{a} u_{b} u_{c} u_{d}+2\left(\delta_{a b} u_{c} u_{d}+\delta_{a c} u_{b} u_{d}+\delta_{a d} u_{b} u_{c}+\delta_{b c} u_{a} u_{d}+\delta_{b d} u_{a} u_{c}+\delta_{c d} u_{a} u_{b}\right)
\end{align}
\section{Fourth moment of displacement in three dimensions (3d)}
\label{app_r4_3d}
\noindent
In this section, we explicitly show the expression of the fourth moment of displacement in three dimensions:

\begin{align}
\begin{aligned}
    \langle\Tilde{\bm{r}}^4\rangle (\tilde t)&= \tilde t^2\left[\frac{\text{Pe}^4 \left(3 \Omega ^4+40 \Omega ^2+240\right)}{9\left(\Omega ^2+4\right)^2}+\text{Pe}^2 \left(\frac{480}{9\left(\Omega ^2+4\right)}+60\right)+60\right]
    \\&
    +\tilde t\frac{\text{Pe}^2}{135 \left(\Omega ^2+4\right)^3}\Bigg[-450 \left(\Omega ^2+4\right) \left(\Omega ^4+48\right)
    \\&-\frac{2 \text{Pe}^2 \left(39 \Omega ^{10}+2203 \Omega ^8+35212 \Omega ^6+235776 \Omega ^4+907200 \Omega ^2+4043520\right)}{\left(\Omega ^2+9\right) \left(\Omega ^2+36\right)}
    \\&
    -\frac{9 e^{-2 \tt} \left(\Omega ^2+4\right)^3 \left(\text{Pe}^2 \left(3 \Omega ^2+32\right)-50 \left(\Omega ^2+16\right)\right)}{\Omega ^2+16}
    \\&
    -\frac{36 e^{-2\tilde t}}{16+\Omega^2} \bigg(\text{Pe}^2 \left(\Omega ^6+64 \Omega ^4-976 \Omega ^2+256\right)+100 (\Omega^4 -16)  \left(\Omega ^2+16\right)\bigg) \cos (\Omega \tt )
    \\&
    -\frac{144 e^{-2\tilde t}}{16+\Omega^2}\Omega  \bigg(\text{Pe}^2 \left(\Omega ^4+108 \Omega ^2-224\right)+100 \left(\Omega ^2+4\right) \left(\Omega ^2+16\right)\bigg) \sin (\Omega \tt )
    \Bigg]
    \\&
    +\frac{\text{Pe}^4}{810 \left(\Omega ^2+4\right)^4 \left(\Omega ^2+9\right)^2 \left(\Omega ^2+36\right)^2}\bigg(43132538880+8017367040 \Omega ^2+4354373376 \Omega ^4
    \\&+1540477440 \Omega ^6+260894256 \Omega ^8+24841136 \Omega ^{10}+1304641 \Omega ^{12}
    +33410 \Omega ^{14}+321 \Omega ^{16}\bigg)
    \\&+ \frac{\text{Pe}^4 e^{-2\tilde t}  \cos (\Omega \tt )}{15 \left(\Omega ^2+4\right)^4 \left(\Omega ^2+16\right)^2}\bigg(\Omega ^{10}-12 \Omega ^8+752 \Omega ^6+4800 \Omega ^4+346112 \Omega ^2-1589248\bigg)
    \\&+\frac{4 \text{Pe}^4 e^{-6 \tilde t}  \sin (\Omega \tt )}{15 \Omega  \left(\Omega ^2+4\right)^4 \left(\Omega ^4+52 \Omega ^2+576\right)^2}\bigg(-\Omega ^2 \left(\Omega ^2+4\right)^4 \left(\Omega ^2+56\right) \left(\Omega ^2+96\right)
    \\&
    +e^{4 \tilde t} \left(\Omega ^2+36\right)^2 \left(\Omega ^{10}-16 \Omega ^8+592 \Omega ^6-3968 \Omega ^4+292864 \Omega ^2-196608\right)\bigg)
    \\&
   + \frac{\text{Pe}^4 e^{-6\tilde t} }{810 \left(\Omega ^2+9\right)^2 \left(\Omega ^2+16\right)^2}\bigg(864 \left(14 \Omega  \left(\Omega ^2-12\right) \sin (2\Omega \tt )+\left(\Omega ^4-73 \Omega ^2+144\right) \cos (2\Omega \tt )\right)
    \\&+\left(3 \Omega ^4+64 \Omega ^2+768\right) \left(\Omega ^2+9\right)^2\bigg)-\frac{2 \text{Pe}^4 e^{-2 \tilde t} \left(3 \Omega ^4+90 \Omega ^2+736\right)}{15 \left(\Omega ^2+16\right)^2}
    \label{r43dtavg}
    \end{aligned}
\end{align}
\end{appendices}




\bibliographystyle{rsc} 
\bibliography{reference} 
\end{document}